# New insights in the lattice dynamics of monolayers, bilayers, and trilayers of $WSe_2$ and unambiguous determination of few-layer-flakes' thickness

**Marta De Luca[1], Xavier Cartoixà[2], Javier Martín-Sánchez[3,4], Miquel López-Suárez[5], Rinaldo Trotta[6], Riccardo Rurali[5], and Ilaria Zardo[1]**

[1] Departement Physik, Universität Basel, 4056 Basel, Switzerland
[2] Departament d'Enginyeria Electrònica, Universitat Autònoma de Barcelona, 08193 Bellaterra, Barcelona, Spain
[3] Departament of Physics, University of Oviedo, 33007 Oviedo, Spain
[4] Center of Research on Nanomaterials and Nanotechnology, CINN (CSIC−Universidad de Oviedo), El Entrego 33940, Spain
[5] Institut de Ciència de Materials de Barcelona (ICMAB–CSIC), Campus de Bellaterra, 08193 Bellaterra, Barcelona, Spain
[6] Departement of Physics, Sapienza University of Rome, P.le A. Moro 2, 00199 Rome, Italy

E-mail: marta.deluca@unibas.ch
ilaria.zardo@unibas.ch

**Abstract**

Among the most common few-layers transition metal dichalcogenides (TMDs), $WSe_2$ is the most challenging material from the lattice dynamics point of view. Indeed, for a long time the main two phonon modes ($A_{1g}$ and $E^1_{2g}$) have been wrongly assigned. In the last few years, these two modes have been properly interpreted, and their quasi-degeneracy in the monolayer has been used for its identification. In this work, we show that this approach has a limited validity and we propose an alternative, more general approach, based on multi-phonon bands. Moreover, we show and interpret all the peaks (about 40) appearing in the Raman spectra of monolayers, bilayers, and trilayers of $WSe_2$ by combining experimental wavelength- and polarization-dependent Raman studies with density-functional theory calculations providing the phonon dispersions, the polarization-resolved first-order Raman spectra, and the one- and two-phonon density of states. This complete study not only offers a method to distinguish between monolayers, bilayers, and trilayers with no need of optical images and atomic force microscopy, but it also sheds light on the interpretation of single and multi-phonon bands appearing in the inelastic light scattering experiments of layered $WSe_2$; some of these bands were never observed before, and some were observed and uncertainly assigned. We promote the full understanding of the lattice dynamics of this material that is crucial for the realization of optoelectronics devices and of novel phononic metamaterials, such as TMDs superlattices.

## 1. Introduction

Atomic-thick layers of semiconducting transition metal dichalcogenides (TMDs) such as $MoS_2$, $MoSe_2$, $WSe_2$, and $WS_2$ display captivating properties and are therefore attracting a growing interest. For example, the direct band gap exhibited by monolayers makes them ideal for optoelectronics and photonics devices [1][2][3], and the valley selectivity of the optical transitions at the direct gap renders monolayers interesting systems in the context of valleytronics [4]. Raman spectroscopy is routinely used as a fast and non-destructive tool to study the properties of TMDs, either for identifying the number of layers in flakes to be employed in devices [1], or for the fundamental investigation of the flakes' lattice dynamics as a function of external perturbations [5][6] and of the exciton-phonon coupling [7][8].

In most of the TMDs, the exact number of layers in a flake can be unambiguously established by optical microscopy [9], photoluminescence spectroscopy [10], or Raman spectroscopy [6]. In the latter, the energy difference between the two high-frequency phonon modes ($A_{1g}$ and $E^1_{2g}$) increases with increasing flakes' thickness and its well-known behaviour has been used to identify the thickness of flakes thinner than ~5-10 layers [1][5][6]. This common practise, well effective for many semiconducting TMDs, has revealed to be quite challenging in $WSe_2$ because of its heavy atoms, which result in a small thickness-dependence of these modes [11]. For this same reason, among the most common TMDs, $WSe_2$ has always been the material whose interpretation of the phononic properties was the most controversial. For example, in all Raman spectroscopy-based works published before 2013, and in one published in 2014 [12], the quasi-degeneracy of $A_{1g}$ and $E^1_{2g}$ modes at ~250 $cm^{-1}$ peculiar of monolayers was unknown, and the broad phonon band at ~260 $cm^{-1}$, nowadays frequently assigned to the simultaneous detection of two longitudinal acoustic phonons (2LA), was attributed to the $A_{1g}$ mode, as reviewed in [8]. Although the correct interpretation of the $A_{1g}$ and $E^1_{2g}$ phonon modes is now widely diffused [6][7][11][13][14][15][16][17][18][19], the late comprehension of the most fundamental phononic properties of $WSe_2$ is likely to be responsible for the several issues still unsolved regarding the interpretation of peaks appearing in the Raman spectra of few-layered $WSe_2$. Hereby, we address all of them: we provide a final assignment to the first-order phonon modes at ~176 and ~310 $cm^{-1}$, we prove that the degeneracy of the $A_{1g}$ and $E^1_{2g}$ modes is not a universal method to identify monolayers as it breaks down for the most common Raman excitation wavelengths (the ones close to 514 nm), we offer a reliable alternative based on the observation of the band at ~260 $cm^{-1}$, we give a new interpretation of 10 of the 17 peaks widely observed in the Raman spectra of $WSe_2$, and we observe and interpret ~25 new peaks.

The need to provide novel, crucial insight in the lattice dynamics of $WSe_2$ is strictly driven by the particular technological interest attracted by this specific material. Indeed, since $WSe_2$ has a relatively small exciton transition energy compared with other TMDs, it is frequently used to form heterostructures with other TMDs [20]; moreover, its huge spin-orbit splitting in the valence band makes it one of the material with the most robust valley polarization [21]. Therefore, it is important to establish the basic physical properties of this material; in particular, a global understanding of the lattice dynamics of thin layers of $WSe_2$ and a correct interpretation of its Raman spectra are fundamental.

Finally, it is worth pointing out that a deep understanding of the phononic properties of thin layers of TMDs is pivotal for allowing the further developments of the recently emerging applications of TMDs, such as the control of their thermal conductivity [22][23][24], the exploitation of interlayer charge transfer dynamics of Van der Waals heterostructures for optoelectronic and photovoltaic applications [25][26], and the investigation of Moiré phonons in twisted layers of TMDs [27] for novel phononic devices.

In this work, we provide a complete and reliable picture of the lattice dynamics of monolayers (1L), bilayers (2L), and trilayers (3L) of $WSe_2$. This is achieved by: *i)* investigating flakes that have good optical quality and are larger than the laser spot; a high spatial resolution ensures that we do not collect Raman signal from flakes having different thickness and that we do not probe edge effects; *ii)* using a Raman setup with very high spectral resolution, which ensures the detection of spectral features that were never observed, and with different excitation wavelengths, such that we are able to achieve resonant and not-resonant conditions of excitation; *iii)* calculating from first-principles the phonon dispersions, the one and two phonon density of states, the Raman tensors of the phonon modes, and the Raman selection rules, which allow us to generate theoretical polarized Raman spectra that can be directly compared to the experiment.

## 2. Results

### *2.1 $WSe_2$ samples*

This work focuses on the Raman investigation of thin-layers of $WSe_2$. In particular, we have measured the 1L, 2L, and 3L samples shown in Figure S1 in the Supporting Information (SI) 1. In some experiments, we have also measured a 4L-thick and a 12L-thick flakes for comparison, as well as the bulk crystal.

The $WSe_2$ flakes were first exfoliated mechanically on a polydimethylsiloxane (PDMS) stamp from a bulk crystal (see section 4.1 in the Methods section). Since the PDMS is optically transparent, the thinnest exfoliated flakes can be identified by optical microscopy in transmission mode. Afterwards, selected flakes are deterministically transferred from the stamp to the target $SiO_2$/Si substrate by using a micro-manipulator. Figure S1 shows the AFM images and profiles on the investigated flakes where their thicknesses can be measured.

### *2.2 Phonon dispersion and phonons' symmetries*

Few-layer TMDs exhibit different phonon symmetries depending on the number of layers. Therefore, basing the interpretation of the Raman spectra of few-layer TMDs on the most-known cases, namely, the bulk and the 1L cases, can be misleading. Ref. [15] provides a comprehensive overview of the symmetry of the first-order Raman-active mode of the 1 to 5 layers-thick $WSe_2$, as well as the calculated phonon dispersion along the ΓM direction in the Brillouin zone (BZ) in the 200-325 $cm^{-1}$ frequency range. The full dispersion along ΓK and MK is also shown for the 1L in references [15][28][29] and for the 2L in ref. [28]. Therefore, we have calculated the full phonon dispersion along ΓM, MK, and ΓK for the 1L, 2L, and 3L by density-functional perturbation theory (DFPT) within the local density approximation, as detailed in the 4.3 paragraph in the Methods. This is shown in SI 2.

In the 1L there are three atoms per unit cell, thus nine phonon branches. The three acoustic branches correspond to the out-of-plane (ZA) vibrations and to the in-plane longitudinal (LA) and transverse (TA) vibrations. The six optical branches correspond to two in-plane longitudinal ($LO_1$ and $LO_2$), two in-plane transverse ($TO_1$ and $TO_2$) vibrations, and two out-of-plane ($ZO_1$ and $ZO_2$) vibrations. For identifying the symmetries that the modes take at Γ, which is given in Fig. S2 (see grey labels), it is necessary to consider the symmetry of the unit cell. The unit cell of the bulk exhibits a $D_{6h}$ point group symmetry and therefore the most known out of plane and in-plane vibrations at Γ are indicated as $A_{1g}$ (~252 $cm^{-1}$) and $E^1_{2g}$ (~248 $cm^{-1}$), respectively [6][15]. Instead, monolayers of TMDs are quasi two-dimensional systems (similarly to graphene), and their vibrational modes belong to the $D_{3h}$ point group. This implies that the labels $A'_1$ and E' have to be used. For the 2, 3, 4, 5, … layers, the group symmetry depends on the parity: TMDs with an odd number of layers exhibit a $D_{3h}$ point-group symmetry, and thus the Raman-active modes belong to the $A'_1$ and E' irreducible representations; an even number of layers corresponds instead to a $D_{3d}$ point group and the modes symmetries are $A^1_{1g}$ and $E^1_g$.

In table S1, S2, and S3 in the SI 3 we also report the computed values of the phonon frequencies at Γ, M, and K, respectively. Since ~35 peaks in the Raman spectra of this material can be attributed to single or multi-phonon processes involving phonon modes at the M and K points of the BZ (as detailed in section 2.4), it is important to know the exact frequency of these phonon modes. Indeed, the approximation of these frequencies for the 1L has introduced an error of ~10 $cm^{-1}$ in the knowledge of the frequencies and contributed to the misinterpretation of many second-order phonon modes. The assignation of these modes is presently reported in comprehensive studies, *e.g.* in refs. [6][8], and therefore largely used.

In the following, we will first discuss peaks in the Raman spectra that can be attributed to single phonon modes at the Γ point (section 2.3), and then focus on the peaks that can be attributed to combination of multiple phonon modes at Γ, M, and K points (and in few cases to single phonon modes at the M and K points) (section 2.4). In this work, we will neglect modes in the low-frequency region (<50 $cm^{-1}$), namely shear and breathing modes typical of layers thicker than 1L and other first-order modes, as they were extensively investigated before, *e.g.*, in refs. [17][30]. Polarization-resolved Raman scattering experiments and calculations were performed in a backscattering geometry in two configurations: xx and xy, being xy the plane containing the flakes and perpendicular to the *c* axis of the bulk (more details are given in section 4.2 in the Methods), as depicted in the top-central sketch in Figure 1.

*2.3 First-order Raman scattering*

Due to energy and momentum conservation laws, first-order Raman scattering can provide access only to phonons at the Γ point. Let us now focus on the optical modes. Based on our calculations, we expect four peaks in a Raman spectrum of $WSe_2$ flakes (~176, ~248, ~250, ~310 $cm^{-1}$). The modes at ~176 and 310 $cm^{-1}$ were poorly investigated in the literature due to their very weak signal, so we start our analysis from those modes.

The peak at ~176 $cm^{-1}$ corresponds to the degenerate $TO_1/LO_1$ modes. By using the calculated Raman tensor for each mode listed in table S1 and following the procedure described in section 4.3 in the Methods, we have obtained the theoretical Raman spectra in Figure 1 (a), left column. Under our scattering configuration, the peak at ~176 $cm^{-1}$ has zero intensity in the 1L (and in the bulk), and it is allowed in the 2L and 3L in both xx (blue lines) and xy (red lines) scattering geometries. It is typical of in-plane vibrations to display the same intensity in xx and xy. In the right column of Fig. 1 (a) we show the experimental Raman spectra, which are in very good agreement with the theory. The frequency of the peak at ~176 $cm^{-1}$ slightly downshifts for increasing layer thickness. The predicted intensity of this mode is quite low (~$10^{-7}$ in our scale), about two orders of magnitude lower than the widely observed modes at ~250 $cm^{-1}$. Indeed, we observe it only with 488 nm (or lower) excitation wavelength, due to resonant Raman conditions, in agreement with refs. [13][17]. However, we disagree with ref. [17] on the interpretation of this mode, as they consider it to be Raman active but forbidden in backscattering for odd-layers $WSe_2$, where its observation is explained invoking the significant portion of the incident and scattered light that is not normal to the sample plane due to the large numerical aperture (NA) of microscope objectives. Instead, we find that this mode is Raman active and forbidden in our geometry (i.e. backscattering, as in the literature) in the 1L, since it requires incoming or scattered light to be polarized along z to be observed, and it is Raman active and allowed in our geometry in the 2L and 3L. In particular, we attribute the ~176 $cm^{-1}$ peak to two degenerate $E^2_g$ modes (at frequency 176.43 $cm^{-1}$ in table S1) in the 2L and to two degenerate $E'^2$ modes (frequency 175.97 $cm^{-1}$ in table S1) in the 3L (not to $E''^2$ modes as suggested in ref. [15]), both allowed in xx and xy backscattering geometries and with similar intensities in the 2L and 3L, which completely explains their observation regardless of the number of layers. The atomic displacements relative to each mode, as derived by our calculations, are also sketched in Figure 1. This interpretation is in agreement with ref. [13], where, however, it is measured/calculated only in the xx configuration.

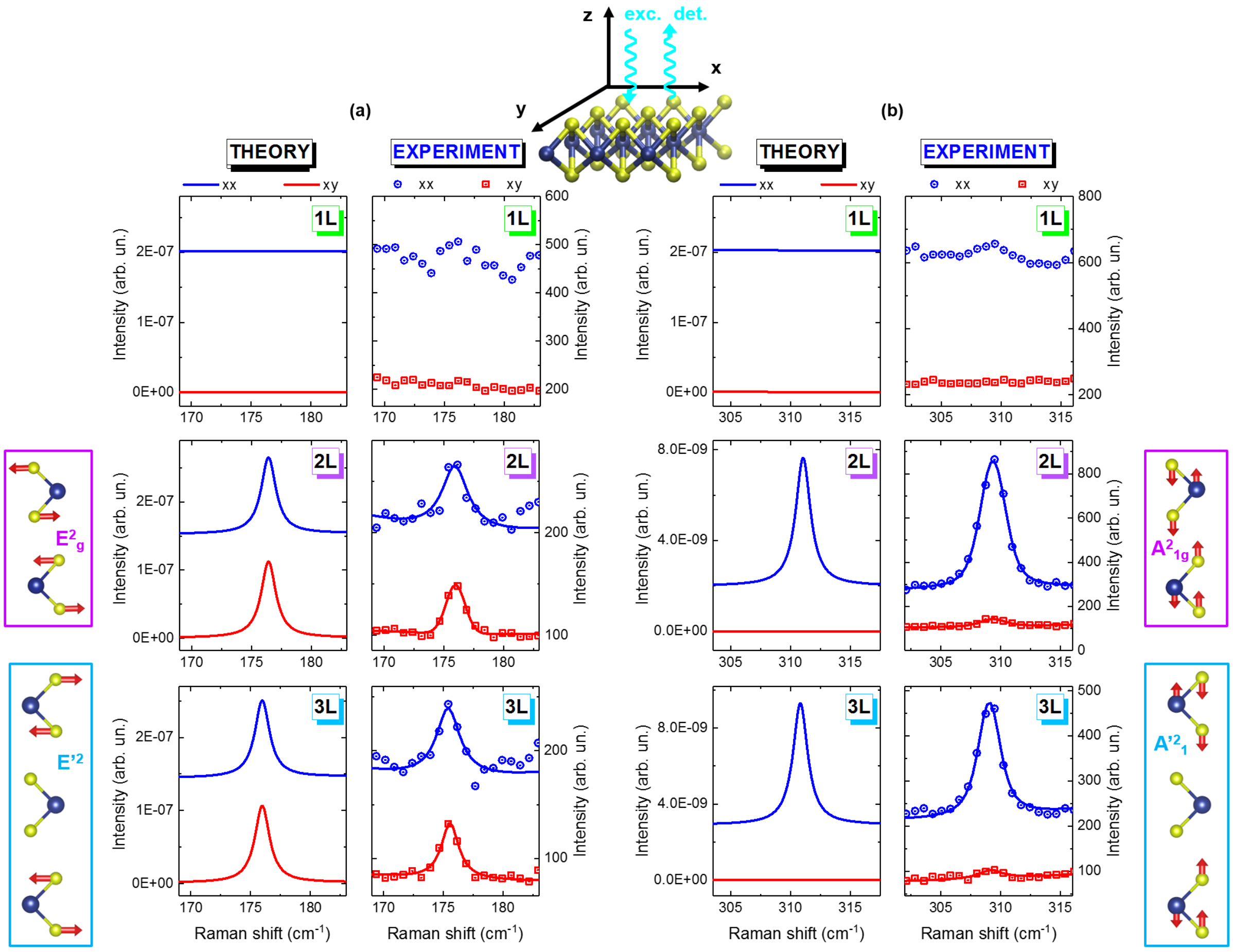

**Figure 1.** (a) Calculated (left column) and measured (right column) Raman spectra on the 1L, 2L, and 3L samples in the xx (blue lines and circles) and xy (red lines and squares) back-scattering geometries (depicted in the schematic at the top-center) magnified in the region of the first-order phonon mode at ~ 176 $cm^{-1}$. Experimental spectra were acquired with the 488 nm excitation wavelength. Solid lines in the experimental spectra are fits to the data. In the experiment, the offset between xx and xy spectra arises from laser background, which is higher for xx configuration (parallel light polarization vectors). In the theory, an offset is added to the xx spectra to facilitate the comparison with the experiment. Atomic displacements for the pertinent phonon modes are depicted in the sketches, where big spheres represent W atoms and small spheres Se atoms. (b) Same as (a) for the first-order phonon mode at ~ 310 $cm^{-1}$.

The peak at ~310 $cm^{-1}$ corresponds to the phonons at Γ of the $ZO_1$ branches. By following the same procedure explained for the 176 $cm^{-1}$ modes, we have obtained the theoretical and experimental Raman spectra reported in Fig. 1 (b), which also in this case are in very good agreement. We find that these modes are Raman inactive for the 1L and Raman active in the 2L and 3L. Similarly to refs. [13][15] and at variance with refs. [8][17], we attribute the peak at ~310 $cm^{-1}$ to a single $A^2_{1g}$ mode in the 2L (at frequency 311.04 $cm^{-1}$ in table S1) and to a single $A'^2_1$ in the 3L (frequency 310.81 $cm^{-1}$). As typical of out-of-plane vibrations, these modes are allowed only in the xx geometry. In the experiments, we observe a small signal from these modes also in the xy geometry due to the high NA of the objective. The frequency of the ~310 $cm^{-1}$ peak downshifts when going from 2L to 3L, in both theory and experiment, as in refs. [8][11]. Similarly to ref. [14], we observe an increase in the intensity of this mode with 514 nm excitation wavelength due to resonant Raman effects.

Both these modes could be used to distinguish the single monolayers from multilayer system, though their intensity, especially for the low-frequency mode, is very low, and this makes this method hard to use unless resonant Raman conditions are exploited [11]. Typically, in order to distinguish the number of layers, the shift between the $A'_1$ and E' modes around 250 $cm^{-1}$ is used – as discussed below.

Figure 2 shows a comparison between the theoretical spectra and experimental spectra magnified in the region of the $A'_1$ and E' modes at ~250 $cm^{-1}$ acquired with different excitation wavelengths, $\lambda_{exc}$ (488, 514, and 633 nm). The first three columns refer to experimental spectra (taken with the indicated $\lambda_{exc}$), and the fourth one displays the theoretical spectra. In the spectra excited with 488 and 633 nm, we observe the same shifts of the phonon modes depending on the number of layers as predicted by the theory. Namely the low-frequency mode downshifts with increasing sample thickness, while the high-frequency mode upshifts (for a quantitative analysis see Figure S3 in SI4). Moreover, the two modes are almost degenerate in the 1L (they correspond to the frequencies of the $TO_2/LO_2$ modes at 250.76 and of the single $ZO_2$ mode at 251.31 $cm^{-1}$ in table S1, this latter being about a factor 4 more intense, for which basically a single peak at ~250 $cm^{-1}$ is observed in xx configuration (while in xy only the mode at 250.76 $cm^{-1}$ is allowed). As shown in table S1, in the 2L there are six modes at ~250 $cm^{-1}$ and in the 3L there are nine. Based on our calculations, in the 2L the low-frequency component (visible with the same intensity in xx and xy in the theory spectra) is assigned to the two degenerate modes at 249.57 $cm^{-1}$, whose symmetry is $E^1_g$, and the high-frequency component (visible only in xx) to a single $A^1_{1g}$ mode with frequency 251.79 $cm^{-1}$. In the 3L, the low-frequency component in the theory spectra in both xx and xy exhibits a splitting, which results in a broadening of the corresponding experimental peak. In agreement with ref. [13] and in disagreement with ref. [15], we attribute the low-frequency component of the xx and xy 3L spectra to a convolution, weighted by their calculated intensity, of four $E'^1$ modes: two modes at 248.57$cm^{-1}$, two modes at 249.60 $cm^{-1}$, and two modes at 249.61$cm^{-1}$. The high frequency component of the xx spectra is attributed to a convolution of two single $A'_1$ modes, at frequency 250.07 and 252.01 $cm^{-1}$, the latter being ~ 4 times more intense than the first, which results in the appearance of a single peak.

The $A'_1$/E' degeneracy in the 1L, along with an increase (decrease) of the high (low)-frequency mode with layer number, is a well-known behaviour typical of $WSe_2$ [15], on which the most common method to identify the number of layers in few layers-thick $WSe_2$ is based [6]. Indeed, in several Raman studies focused on the fundamental properties of TMDs or integration of monolayers into devices, whenever a single peak at ~250 $cm^{-1}$ is observed, the presence of a monolayer is claimed; see, *e.g.*, refs. [31][32][33][34]. Here, we stress that this is the case only for $\lambda_{exc}$=488 and 633 nm, while for the $\lambda_{exc}$=514 nm (as well as for $\lambda_{exc}$=521 nm, not shown here), the quasi degeneracy almost takes place also in the 2L. We stress that the spectra in xx and xy configurations are indeed very similar, and with a lower-resolution, more conventional, spectrometer would appear indistinguishable. Also the 2L and the 3L would likely be indistinguishable. In this respect, we stress that in ref. [8] the $A'_1$ and E' modes in the 2L and even thicker layers are indeed not distinguishable with the 532 nm excitation wavelength. As a consequence of these observations, we suggest that the most used method to identify monolayers has a limited validity and its reliability depends on the excitation wavelength. Besides this, we stress that a method based on the distinction between very close peaks can be applied only when setups with high spectral resolution are available (at least below 3 $cm^{-1}$): if this resolution is not available, the spectra taken any excitation wavelength on the 1L and the 2L samples would both display a single peak at ~250 $cm^{-1}$ and the two samples would have been both identified as monolayers. In most of the TMDs, a high spectral resolution is not necessary, as the $A'_1$ and E' modes are quite far in frequency (~17 $cm^{-1}$ in $MoS_2$ and 65 $cm^{-1}$ in $WS_2$ [6]); however, we have just shown that applying the method developed for $MoS_2$, $WS_2$, *etc.* to a material as $WSe_2$ can be misleading. As a matter of fact, the peculiarity of $WSe_2$ is the reason why for a few years the $A'_1$ and E' modes in $WSe_2$ were not properly identified, and what is now known as 2LA band at ~260 $cm^{-1}$ was interpreted as the $A'_1$ mode until 2013/2014. In a recent work [10], a single broad peak at ~250 $cm^{-1}$ attributed to degenerate $A'_1$/E' modes is observed from 1 to 6 layers, and indeed its weak upshift with increasing layer number is considered not applicable to identify layer thickness.

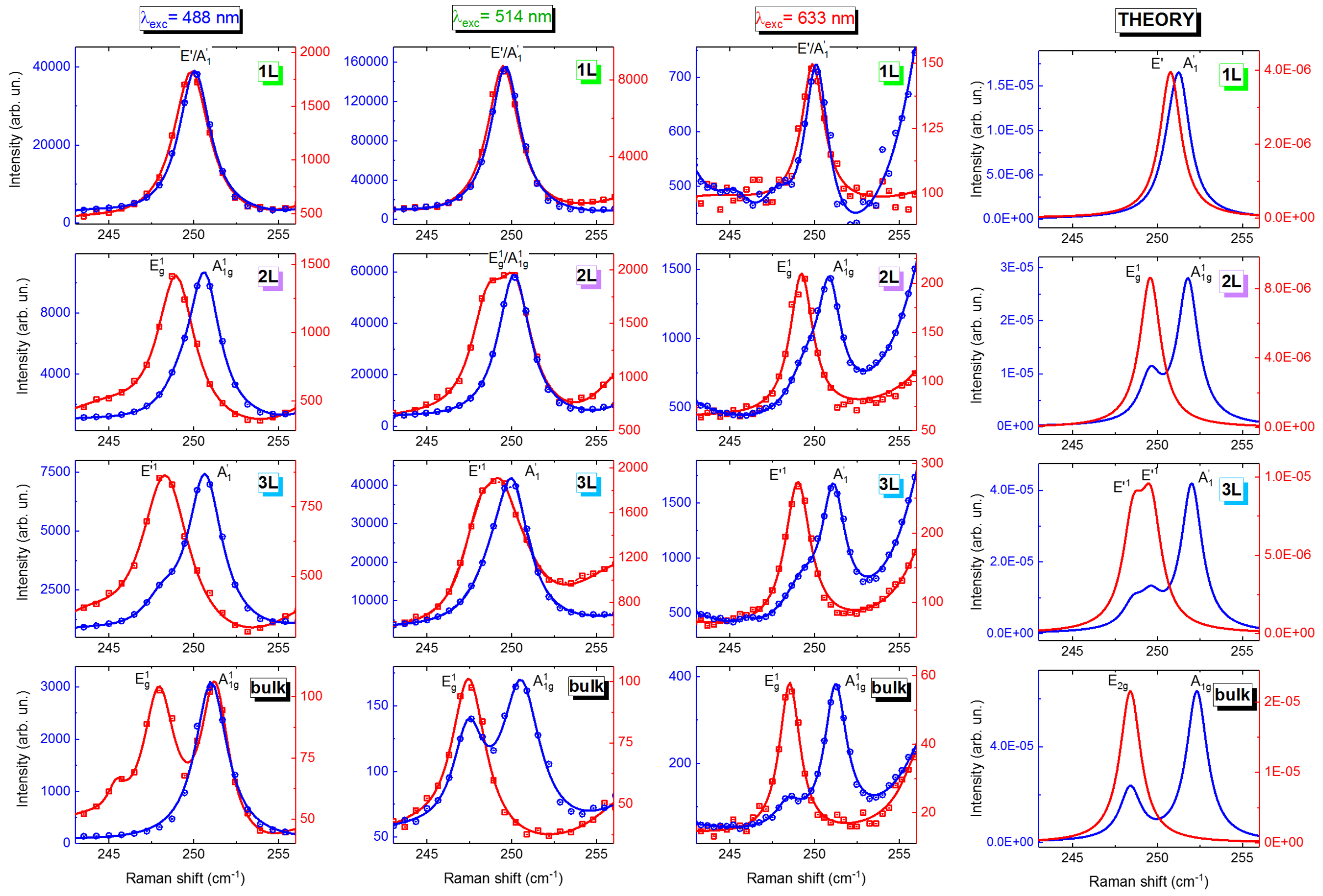

**Figure 2.** Experimental (first three columns) and calculated (fourth column) Raman spectra of the 1L, 2L, 3L, and bulk in the xx (blue lines and circles) and xy (red lines and squares) scattering geometries magnified in the region of the first-order phonon modes at ~ 250 $cm^{-1}$. Experimental spectra were acquired with the 488, 514, and 633 nm excitation wavelengths. Solid lines in the experimental spectra are fits to the data. For each couple of spectra (xx and xy), intensity scales were chosen in order to approximately align the background signal as well as the maxima, to facilitate the comparison between the two polarized spectra and also with the theory.

Let us now focus on the intensity of the peaks, which mostly depends on Raman selection rules. Regardless of the layer thickness, the theory predicts that while in the xx configuration both modes can be observed, although with different intensities, in the xy only the low-frequency modes are allowed. We notice that the Raman spectra of the 1L, 2L, and 3L samples collected with $\lambda_{exc}$=488 nm and 633 nm are in good agreement with this prediction. Small deviation from the theory can be accounted for by the use of high NA objectives and by the effect of small errors in the polarizer angles (this issue is discussed in Figure S4 in SI5). These two effects are quite small and can be responsible for a partial relaxation of selection rules, but cannot render a forbidden mode equally visible or more visible than an allowed mode. Therefore, the presence of a low-intensity $E^1_g$ mode (~950 counts over the background, after quantitative analysis) and of a high-intensity $A^1_{1g}$ mode (~1100 counts) in the 2L measured with 514 nm in xy geometry (notice the similar behaviour, though less striking, in the 3L) cannot be explained by experimental effects alone, but only by resonant Raman effects, which are likely to be responsible also of the fact that the bulk sample does not obey the selection rules when measured with 488 nm. Understanding the behaviour of the 2L with 514 nm is important, as it is responsible for the apparent modes' degeneracy (discussed above) that impedes from distinguishing 1L and 2L with $\lambda_{exc}$= 514 and 521 nm. In a recent resonant Raman study, it was observed a strong enhancement of the $A^1_{1g}$ mode in the 2L in the 510-570 nm excitation range due the resonance with the energy of the A' exciton [17]. This effect, combined with the experimental effects above discussed, explains why in the 2L it is the most intense peak even in the xy configuration where it should be forbidden. The resonant behaviour of the $A^1_{1g}$ mode is corroborated by the one order of magnitude increase of its intensity compared to the spectra collected with other excitation wavelengths. For the bulk sample, there is a good agreement between theory and experiment, with the exception of the spectra taken with 488 nm, where the $A^1_{1g}$ mode is likely to be resonant and therefore it dominates the spectra taken in xx geometry and it is well visible also in the xy geometry.

The observations reported in Figure 2 demonstrate that the shift and intensity of $A'_1$ and E' modes is not a reliable and universal method to identify the thickness of thin $WSe_2$ flakes. Therefore, we have developed a new procedure, described in the next section.

### *2.4 Second-order Raman scattering*

Besides the four peaks in the Raman spectra of thin $WSe_2$ flakes originating from first-order vibrational processes (*e.g.* the emission or absorption of one phonon), there are several other peaks in the spectra. They are mostly attributed to multiple-order Raman scattering, with at least two phonons involved. Due to momentum conservation laws, it is possible to combine two (or more) phonons having zero momentum (*i. e.* phonons at Γ), or pairs of phonons with equal and opposite momentum (plus, in general, any combination of phonons whose total momentum is zero). High-order Raman processes are less probable than first-order processes, and therefore weaker, but they can be resonantly enhanced when the excitation wavelength matches some interband transitions of the semiconductor. Resonant Raman experiments can indeed provide important information on the band structure. Besides that, this enhancement [6], or other effects, such as the presence of defects [35], or nanocrystals [16], can make it possible to detect single phonons at the edge of the BZ (*e.g.*, M and K points), thus apparently bypassing the limitations of the momentum conservation law in one-phonon processes (when single phonons at BZ boundaries are activated by a defect, the total momentum is actually conserved, owing to the momentum contribution of the defect). In the following, we will divide the discussion of all these processes in two parts. In Figure 3, we focus on the broad band at ~260 $cm^{-1}$, which provides us with a new, powerful method to identify the number of layers in few-layered $WSe_2$. In Figure 4, on the other hand, we focus on all the other peaks. In table 1, we then assign all the peaks observed in Figure 3 and 4 to single or multiple scattering processes.

Figure 3 displays a close-up of the Raman spectra in the 252-267 $cm^{-1}$ range (commonly labelled as 2LA band [6][11][13][15][17]) taken with $\lambda_{exc}$=488, 514, and 633 nm (from left to right), on the 1L, 2L, and 3L samples in the xx scattering geometry. Each spectrum was normalized to its maximum (which is indicated by the corresponding arrows) in order to highlight that there is a general wavelength-independent behaviour, consisting in a gradual shift of the maximum towards lower frequencies increasing the number of layers. A similar behaviour was observed in some previous works, *e. g.* [13][14][17][36], but it was not commented. Here, we show that this spectral signature allows for a clear distinction between 1L, 2L, and 3L samples. The spectra in the xy geometry, along with those of the bulk sample, all plotted with their own intensity scales, can be found in Figure S5 in the SI6. The interpretation of this broad Raman band in terms of single phonons at the edge of the BZ and second-order processes for the 1L is discussed in detail in table 1 and for the 1L, 2L, and 3L in Figures S6 and S7 in SI7. A proper interpretation of the so-called 2LA band in bulk and monolayers of $MoS_2$ was attained in ref. [37], but due to the lack of the investigation of the 2L and 3L, it is not clear whether the 2LA band could be used identify layer thickness. In $WS_2$, the dependence of the intensity of the 2LA band on layer thickness was proposed as a method to distinguish 1L, 2L, and 3L [38]. The method clearly works for $\lambda_{exc}$=488 nm, while the dependence of the intensity on layer thickness is either weak or reversed for other excitation wavelengths. Here, we stress that the method we just described to identify few layers-thick flakes satisfies four important requirements: *i)* validity independent of the excitation wavelengths (in the blue-to-red visible range typically employed for Raman spectroscopy); *ii)* applicability in conventional Raman setups (*i. e.*, necessary spectral resolution $\geq$ 5 $cm^{-1}$); *iii)* no necessity of polarization-dependent measurements; iv) independency of the intensity of the peaks, namely independency of the efficiency of the setup as well as on the resonance conditions (flakes with different thickness display resonances for different wavelengths, therefore any method based on the intensity of the modes and not on their frequency is not universally applicable).

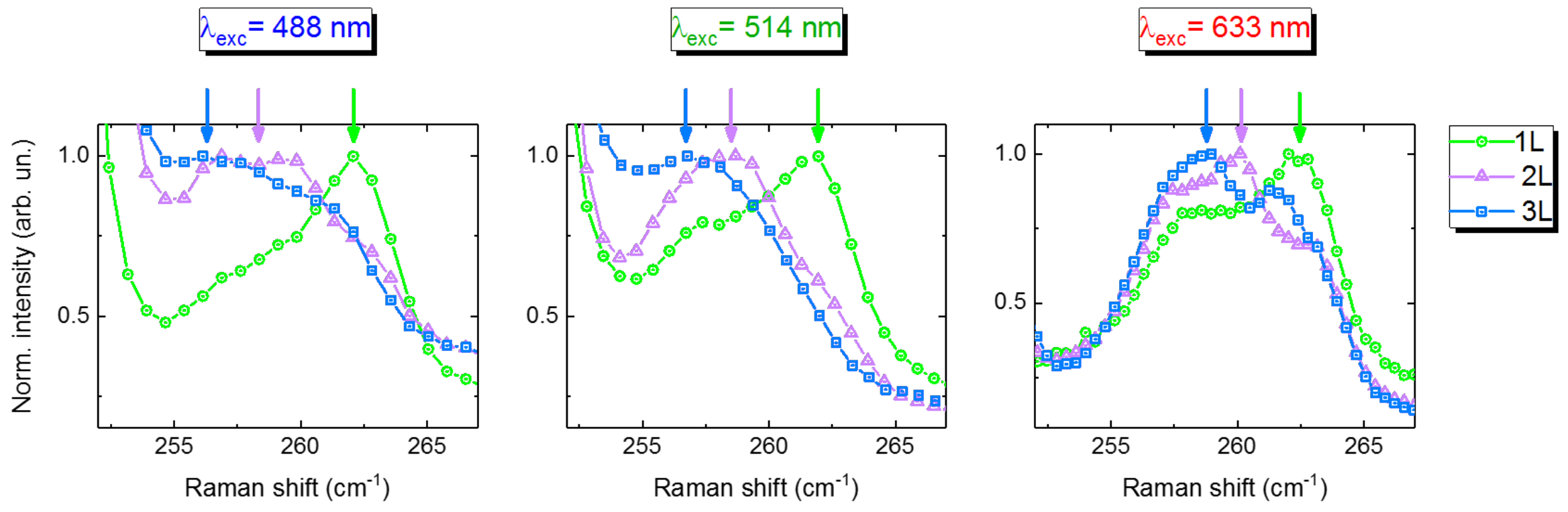

**Figure 3.** Raman spectra of the 1L (green line+circles), 2L (purple line+triangles), and 3L (blue line+squares) samples in the xx scattering geometry magnified in the region of the phonon modes at ~ 260 $cm^{-1}$. Spectra were collected with the 488, 514, and 633 nm excitation wavelengths (from left to right columns, respectively). Intensities are normalized to the maxima (absolute intensities can be found in Fig. S5). The arrows indicate approximately the maximum of each spectrum and are a guide to the eye to identify the downshift in the spectral weight when going from 1L to 3L.

Figure 3 displays the whole Raman spectra of the 1L in the 55-560 $cm^{-1}$ range, divided into four spectral regions for clarity reasons. Spectra were acquired with $\lambda_{exc}$=488 nm in the xx (first row) and xy (second row) scattering geometries. Measurements taken with 514 and 633 nm exhibit similar spectral features, in some cases with different intensities due to resonant Raman effects and we will discuss the differences in detail whenever they are relevant. Open symbols are experimental data, thin lines are the single lorentzian contributions, and thick lines are the cumulative fits. The experimental frequencies obtained by these fits are listed in the first column of table 1 (if visible in both xx and xy geometries, the frequencies were averaged). The previously discussed peaks at ~176 and 310 $cm^{-1}$ are absent as they are forbidden in the 1L, as shown in Fig. 1. The weak/broad peaks indicated by dashed grey lines in Fig. 4 were not considered in table 1, unless they were clearly detectable with other excitation wavelengths. For example, due to a huge electronic resonance occurring for the $A'_1$/E' modes of the 1L with the 488 (and 514) nm excitation wavelengths [13][17], in the 488 nm spectra in Fig. 4 the peak at 250 $cm^{-1}$ is much more intense than the peaks in the 239-267 $cm^{-1}$ range, which makes the deconvolution of these latter not fully reliable. As those peaks were instead very clearly detected with the 633 nm wavelength (see Figure S8 in the SI8), in table 1 we have reported their frequencies. Indeed, the 633 nm wavelength is non-resonant for the $A'_1$/E' modes in the 1L [13], which makes it ideal to study the modes in the ~239-267 $cm^{-1}$ range.

Besides the first-order $A'_1$/E' modes at ~250 $cm^{-1}$ extensively discussed, in table 1 all the other phonon modes are also assigned (assignations are given in the third column) to sum or differences of two phonons with total null momentum (overtones or combinations of two different phonons) or to single phonons at the BZ edges (these latter are marked with a *). The resulting theoretical frequencies are reported in the second column. Within the experimental and theoretical errors (discussed in section 4.3), we obtain a very good agreement between theory and experiment for all the peaks. For the assignation in terms of single phonons at the BZ edges we have considered the frequencies reported in tables S2 and S3 in SI3. Instead, for the assignation in terms of sum and difference of phonons at Γ, K, and M, our starting point was a calculation of the two-phonon density of states (2DOS), which is shown in Figures S9 in SI9 (for the 1L, 2L, and 3L). Namely, we have determined the combination of phonons that were corresponding to peaks in the 2DOS and looked for correspondences with the experimental frequencies. As the 2DOS is known to not be a direct simulation of the second-order Raman spectra, especially in presence of resonance effects [37], we have completed the experiment-theory comparison by also considering all the combinations of modes that did not result into clear peaks of the 2DOS but that clearly matched experimental frequencies. In this way, we could finalize the interpretation of all the peaks. As detailed in the table, we suggest that the 2LA band of the 1L is not only composed by two LA phonons at M, but also by single BZ-edge phonons: $ZO_1$(K), $LO_2$(K), $ZO_2$(K), and $ZO_1$(M), among which only this latter has been considered in the literature [6]. This attribution is corroborated by the observation that in the 2DOS, shown in Figure S9, no clear peak is distinguishable at ~260 $cm^{-1}$. The deconvolution in terms of the five contributions for the 1L is shown in Figure S8 in the SI8. For the 2L and 3L samples, there are even more contributions, see Figures S6 and S7 in SI7. A systematic Resonant Raman study of the 2LA band would further shed light on the origin of the different peaks contributing to the 2LA band [39].

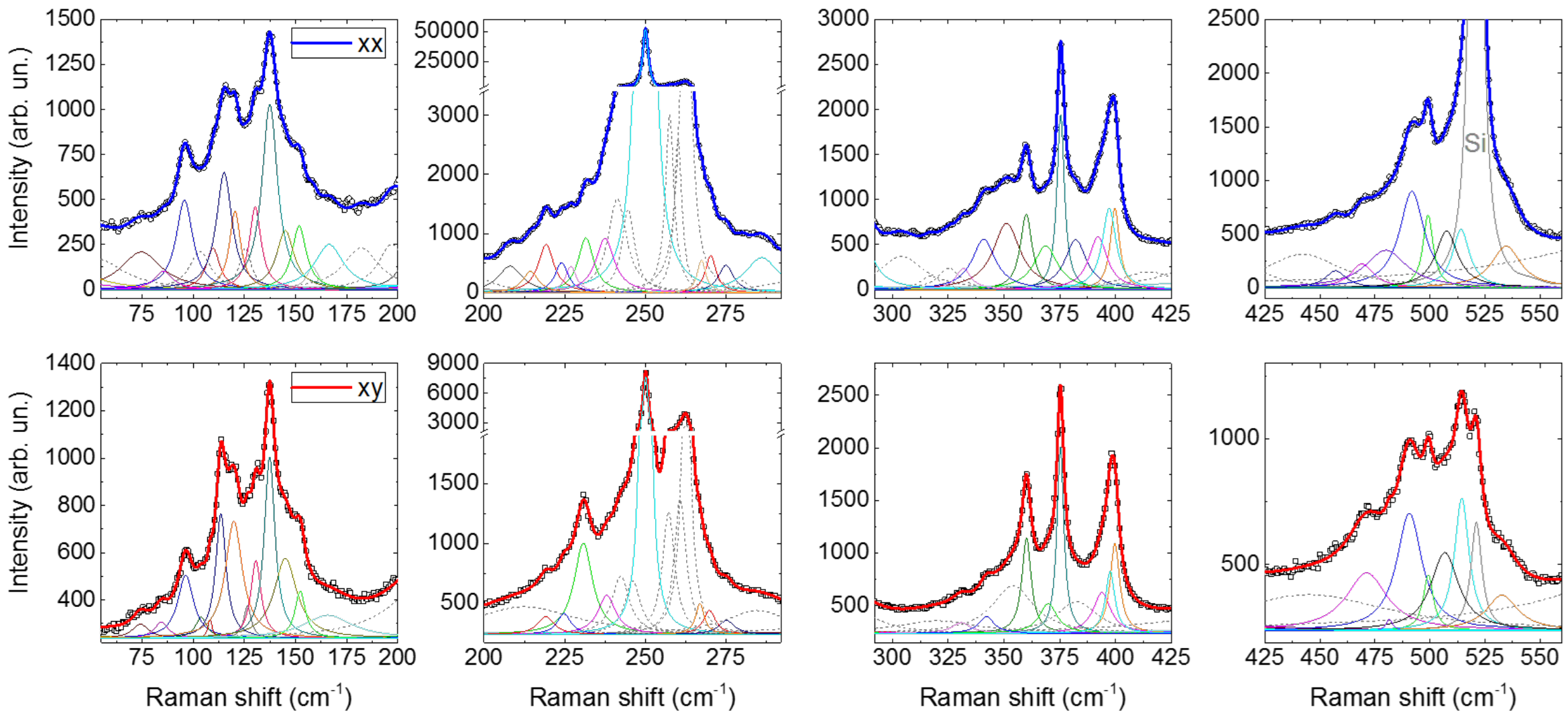

**Figure 4.** Experimental Raman spectra acquired with the 488 nm wavelength on the 1L sample in the xx (blue lines and circles) and xy (red lines and squares) scattering geometries divided into four different frequency ranges for clarity. Thick solid lines are fits to the data, thin colored lines are the single lorentzian components (a same color in xx and xy spectra indicate a same mode). The frequencies extracted by the fits are given in table 1. Dashed grey lines refer to lorentzian curves that do not correspond to clear peaks in the experimental spectra and therefore have not been included in table 1, unless they were more clearly detectable with other excitation wavelengths.

We stress that 17 of the peaks listed in the table were reported and interpreted in previous works [8][16] ([6] is a review work based on [11][15]). We partially agree with the interpretation of seven (~116, 219, 231, 245, 261, 374 and 398 $cm^{-1}$) of the 17 peaks. At a first sight, our assignment of these seven modes looks in disagreement with refs. [6][8][16], just because we use a different nomenclature: for the modes at M and K, whose symmetry is not known, we prefer not to use the symmetry that the modes take at Γ, we rather use the label of the branch indicated in Figure S2 in SI2. For the other 10 modes, instead, there is real disagreement. Finally, there are two more works [10][14] reporting three peaks that we observe and that are not discussed in refs. [6][8][16]: 225, 242, and 360 $cm^{-1}$. We have reassigned also these three peaks. All the other peaks (~25) listed in table 1 were never reported before. In Raman spectra of small nanocrystals at low temperature some more peaks compared to table 1 are observed due to phonon confinement and mostly attributed to third order processes [16]. Moreover, we stress that the contribution of single or multiple phonon modes at K was never considered in the literature, with the exception of ref. [13] for the assignment of the 225 $cm^{-1}$ peak to a E(K) mode, with which we disagree as just discussed (according to our calculations, this mode, labelled as $LO_1$ in table S3, has a frequency of 216 $cm^{-1}$). Finally, at variance with refs. [6][8][16], we do not assign peaks to third order processes (like 3TA (M) and 3LA (M)), because they do not match any experimental frequency in table 1 and because they are unlikely to give detectable Raman signal (both the violation of momentum conservation and the high number of involved phonons would make the associated signal very weak).

By following the same procedure described for the 1L, it is possible to carry out a similar investigation also for the 2L and the 3L. Therein, however, there are many more phononic branches, so many more possible phonon combinations resulting in almost the same frequency. Including so many modes in a fitting procedure would make the analysis somewhat arbitrary. In Figure S6 and S7 we provide an example of a possible procedure for the broad 2LA band and in Figure S10 in the SI10 for the peaks at 220, 224, and 242 $cm^{-1}$.

| **1L** | | |
|---|---|---|
| **Exp. freq.** | **Theor. freq.** | **Mode assignation** |
| 74.4 | 76.3 | $ZO_2$(M) - $TO_1$(M) |
| 85.0 | 85.9 | $TO_1$(K) - ZA(K) |
| 96.0 | 94.2<br>95.3 | $LO_1$(M) - ZA(M)<br>$TO_1$(M) - TA(M) |
| 108.9 | 107.1 | $TO_2$(M) - ZA(M) |
| 114.3 | 114.5<br>114.9<br>115.1<br>116.3<br>116.5 | $TO_2$(K) - TA(K)<br>$ZO_1$(K) - LA(K)<br>$LO_2$(M) - LA(M)<br>$LO_1$(K) - TA(K)<br>$LO_2$(K) - LA(K) |
| 120.1 | 119.5 | $ZO_2$(K) - LA(K) |
| 130.5 | 130.7<br>130.1 | $TO_2$(M) - TA(M)<br>*LA(M) |
| 137.4 | 135.2<br>136.2 | $LO_2$(K) - ZA(K)<br>$ZO_1(\Gamma)$ - TO/$LO_1(\Gamma)$ |
| 145.0 | 146.8 | $LO_2$(M) - TA(M) |
| 152.0 | 151.8 | other BZ points |
| 158.4 | 157.4<br>159.0 | $ZO_1$(K) - TA(K)<br>$LO_2$(K) - TA(K) |
| 208.1 | 209.6 | *$TO_1$(K) |
| 219.2 | 220.4 | TA(M) + ZA(M) |
| 224.4 | 223.6 | TA(K) + ZA(K) |
| 231.2 | 228.5<br>229.1 | TA(M) + LA(M)<br>*$TO_2$(M) |
| 237.8 | | |
| 239.9 | 242.3 | TA(K) + LA(K) |
| 244.2 | 244.0<br>245.2 | 2 x ZA(M)<br>*$LO_2$(M) |
| 249.9 | 250.8<br>251.3 | $TO_2$/$LO_2$ $(\Gamma)$: E'<br>$ZO_2$ $(\Gamma)$: $A'_1$ |
| 258.2 | 257.3<br>258.9 | *$ZO_1$(K)<br>*$LO_2$(K) |
| 261.0 | 260.2 | 2 x LA(M) |
| 262.3 | 261.9 | *$ZO_2$(K) |
| 263.3 | 263.3 | *$ZO_1$(M) |
| 266.6 | 266.1 | ZA(K) + LA(K) |
| 270.1 | 270.0 | *$ZO_2$(M) |
| 318.6 | 315.7<br>316.1 | $TO_1$(M) + ZA(M)<br>$LO_1$(K) + TA(K) |
| 331.3 | 331.0 | other BZ points |
| 341.5 | 339.9 | $TO_2$(M) + ZA(M) |
| 351.2 | 352.3 | 2 x TO/$LO_1(\Gamma)$ |
| 360.0 | 361.7<br>361.8 | $ZO_1$(M) + TA(M)<br>$ZO_2$(K) + TA(K) |
| 369.3 | 367.2<br>368.4<br>369.0 | $LO_2$(M) + ZA(M)<br>$ZO_2$(M) + TA(M)<br>other BZ points |
| 375.4 | 373.3 | $LO_2(M) + LA(M)$ |
| 382.4 | 381.0 | $ZO_1(K) + ZA(K)$ |
| 393.0 | 392.0 | $ZO_2(M) + ZA(M)$ |
| 399.7 | 399.7<br>400.1 | $ZO_1(K) + LA(K)$<br>$ZO_2(M) + LA(M)$ |
| 480.8 | 479.5 | $LO_1(M) + ZO_1(M)$ |
| 491.3 | 492.4 | $ZO_1(M) + TO_2(M)$ |
| 499.2 | 499.1 | $ZO_2(M) + TO_2(M)$ |
| 507.1 | 508.5 | $ZO_1(M) + LO_2(M)$ |
| 514.4 | 514.6<br>516.2 | 2 x $ZO_1(K)$<br>$ZO_1(K) + LO_2(K)$ |
| 533.8 | 533.3 | $ZO_1(M) + ZO_2(M)$ |

**Table 1.** Interpretation of the peaks observed in the Raman spectra of the 1L. First column: experimental frequency derived from fitting the Raman spectra displayed in Figure 4 taken with $\lambda_{exc}$=488 nm (the frequencies of the modes observed in both xx and xy configurations have been averaged). The error on the frequencies is ~ ±1 $cm^{-1}$. Second column: calculated frequencies based on the tables S1, S2, and S3 in the SI (at $T$= 0 K). Third column: combination of phonon modes matching the experimentally observed peaks; this combination results in the frequencies reported in the second column. In some cases, more than one combination of phonons is compatible with the experimental frequency and therefore all the compatible combinations are reported. Modes marked with * are compatible with first-order phonon modes at BZ boundaries, which are indicated. The label 'other BZ points' indicate peaks that result from the combination of modes that are not at the three high symmetry points in the BZ and whose notable contribution to the Raman spectrum is suggested by our calculation of the 2DOS (see SI9). As the peak at 237.8 $cm^{-1}$ does not correspond to any calculated frequency, it might be due to a third- or fourth-order process. For the well-known first order modes at ~250 $cm^{-1}$, also the symmetry at Γ is given. The peaks in the range 239-267 $cm^{-1}$, despite being visible also in the spectra in Figure 4, were more clearly detected with 633 nm wavelength (see Figure S7 in the SI7), and therefore for those peaks the values in the table refer to spectra collected with 633 nm.

In table 1 we did not take into account differences between results obtained in xx and xy geometries because the calculation of symmetries of all the modes at M and K points and the calculation of the selection rules for second or higher order modes is beyond the scope of this article. Such thorough analysis is provided *e.g.* in ref. [37] for $MoS_2$, and in ref. [40] for the modes at Γ in bulk $WSe_2$. However, we could assign all the peaks even without applying selection rules. We believe that selection rules would be helpful for the cases where there is more than one possible interpretation of the peaks because they would allow to rule out some of the assignations. For all the other cases, the assignation is univocal as there was no other possible single or multiple phonon combination matching the experimental frequencies.

For most of the frequencies in table 1, we cannot say if the high-order modes (or single modes at BZ edge) that should be very weak (or forbidden) are instead observable because of resonant Raman effects, because we observe most of them with all the wavelengths (though with different relative intensities), because the resonant Raman profile is known only for very few of these modes [7][8][13][17], and because the few available results are not fully in agreement with each other [7][13][17]. Our results suggest that it is not necessary to exactly match the resonance conditions to be able to observe (and interpret) these peaks.

## 3. Conclusions

We have presented a complete investigation of the phononic properties of $WSe_2$ flakes, with focus on monolayers, bilayers, and trilayers. A combination of Raman spectroscopy experiments (performed in- and off-resonance depending on the purpose) with DFT calculations has allowed us to:

*i)* address the open issues in the interpretation of first-order phonon modes by clarifying the symmetry of the modes. In particular, we have explained the independence from layer parity of the ~176 $cm^{-1}$ peak by attributing it to a $E^2{}_g$ mode in the 2L and to a $E'^2$ mode in the 3L;

*ii)* demonstrate that the observation of the degeneracy of the first-order $A'_1/E'$ modes at ~250 $cm^{-1}$ in monolayers is *de facto* wavelength-dependent and, therefore, it is not a reliable and universal method to identify the thickness of thin $WSe_2$ flakes. Instead, the multiple phonon band at ~260 $cm^{-1}$ undergoes a clear shift of the maximum towards the low frequencies when going from the 1L to the 3L samples, regardless of the excitation wavelength in the visible range. Therefore, that band is an unambiguous Raman fingerprint for identifying single- and few-layered $WSe_2$ flakes;

*iii)* observe ~25 new peaks in the Raman spectra of the monolayers and provide a complete interpretation in terms of combinations of two phonons at the BZ centre and boundaries or of single phonons at the BZ boundary. We show how to apply this type of investigation to bilayers and thicker flakes. We also provide a new interpretation of 10 among the 17 peaks previously observed in the Raman spectra in the literature, including the 2LA band.

## 4. Methods

### *4.1 Preparation of samples and AFM characterization*

The samples were prepared by mechanical exfoliation followed by the dry-transfer of the $WSe_2$ flakes on a PDMS stamp [41]. Our $WSe_2$ bulk crystals were purchased from hqgraphene. The crystal was first exfoliated on the stamp. The exfoliated flakes are there localized in a conventional microscope by measuring optical transmission (the PDMS stamp is transparent). Once the flakes are localized, they can be transferred by using a micro-manipulator on any substrate (in our study it was Si with 90 nm of $SiO_2$ on top) with micrometer resolution in the positioning. In this way we have obtained the investigated 1L, 2L, 3L, 4L, and 12(±1)L flakes. We have also measured a reference bulk sample, from which the flakes were exfoliated. For $WSe_2$ the shifts of phonon modes is almost negligible in samples thicker than 7-8 layers [10]; indeed, we found that the 12L flake displays almost the same Raman spectra of the bulk. All the flakes are considerably larger than the diffraction-limited laser spot, and this made the spectroscopic measurements easy and reliable. Indeed, this technique allows to obtain a higher monolayer yield and larger monolayers compared to conventional scotch-tape exfoliation. Moreover, the position of the flakes on the desired substrate is deterministic. All these characteristics render this technique interesting for the inclusion of TMDs flakes in devices. Besides those advantages, the other advantage is that the optical quality is higher than the one of the flakes exfoliated with scotch tape. For comparison, we measured by Raman spectroscopy also flakes obtained by conventional exfoliation based on scotch tape, and no relevant differences were detected.

The AFM characterization was performed in tapping mode using an AFM Neaspec set-up in ambient conditions. The processing and analysis of the experimental data was realized employing the WSxM software [42].

### *4.2 Raman experimental details*

Raman measurements were performed in backscattering geometry on $WSe_2$ flakes at room temperature, with excitation wavelength of 632.8 nm provided by a HeNe laser and of 488, 514, and 521 nm provided by a $Ar^+Kr^+$ laser. The power was 0.1 mW (selected after checking that no heating effects were induced). The laser beam was focused with a 100x objective (numerical aperture of 0.80) and the measured spot size is basically diffraction-limited. The scattered light was collected by a T64000 triple spectrometer (Horiba) equipped with 1800 g/mm gratings and liquid-nitrogen cooled multichannel charge couple device detector. Polarization-resolved Raman scattering experiments and calculations were performed in backscattering geometry: the incident photon wavevector ($k_i$) is antiparallel to the z axis and the scattered photon wavevector ($k_s$) is parallel to z (=001). As a consequence, the light polarization vectors, $\varepsilon_i$ and $\varepsilon_s$, lie in the xy plane (x=100, y=010), which is the plane of the sample, mounted on a micrometric stage; $\varepsilon_i$ and $\varepsilon_s$ can be separately controlled to obtain the desired scattering geometries. Spectra were collected (and calculated) by selecting scattered light polarized either parallel or perpendicular to the polarization direction (x) of incident light, which in the so-called Porto notation correspond to -z(xx)z and -z(xy)z, respectively, in the text indicated as xx and xy for simplicity. For each flake at least two points were investigated and no differences were found. All the displayed Raman spectra were acquired with a 120s acquisition time and they were averaged 3-5 times.

### *4.3 Theoretical calculations*

We have calculated the ground-state geometry, the electronic structure and the Raman spectra of thin layers and bulk $WSe_2$. In the case of multilayers we have considered the 2H structure, which is the most stable polytype. We have used the ABINIT code [43], optimized the lattice vectors and the atomic position using optimized norm-conserving pseudopotentials [44] with a plane-wave cutoff of 41 Ry and a 16x16x1 k-point grid. We have used density functional perturbation theory (DFPT) to compute the phonon dispersion relations and the Raman susceptibilities. For the bulk we have used 4 k-points along $k_z$. The exchange-correlation energy has been calculated within the local density approximation (LDA) in the Ceperley-Alder parameterization. No van der Waals corrections have been included. It has previously argued that this computational setup yields frequencies in better agreement with experimental data [30].

Frequencies estimated by DFPT are at 0 K, while all the experimental ones are measured at room temperature. However, we expect this not to be a serious limitation of the predictive nature of the theory results for several reasons: $WSe_2$ is one of the materials with the weakest temperature-dependence of the A'$_1$/E' and 2LA modes (~1 $cm^{-1}$ [12]); the $T$-dependence of all the other modes is not known; experimental frequencies have a ±1 $cm^{-1}$ uncertainty (due to the fitting more than to experimental errors).

In order to properly account for the scattering geometry used in the experiment, the Raman intensity of each mode $n$ has been calculated as

$$I_n \propto |\varepsilon_i \, R_n \, \varepsilon_s|^2$$

where $R_n$ is the Raman susceptibility tensor calculated *ab initio* in the reference system of the experiment, while $\varepsilon_i$ and $\varepsilon_s$ are the polarization vectors of the incident and scattered light, respectively. This method is deeply explained in references [45][46]. Once the intensity for each phonon mode has been calculated, Raman spectra are generated by summing up Lorentzian functions, each associated to a calculated mode frequency. The full width at half maximum used for the Lorentzian functions was 1.5 $cm^{-1}$, slightly smaller than the experimental one in order to render visible possible modes' splitting. The intensities in the experimental spectra are directly comparable with each other, and so do the theoretical spectra with each other; however, experimental and theoretical intensities cannot be compared because different 'arbitrary units' are employed. Moreover, the relative intensity of spectra taken on samples with different number of layers is not meaningful because different number of layers result in different resonant Raman conditions involving excitonic effects [6][47], which are not taken into account in our calculations.

## Acknowledgements

This project has received funding from the Swiss National Science Foundation research grant (Project Grant No. 200021_165784). M. D. L. acknowledges support from the Swiss National Science Foundation Ambizione grant (Grant No. PZ00P2_179801). R.R. acknowledges financial support by the Ministerio de Economía, Industria y Competitividad (MINECO) under grant FEDER-MAT2017-90024-P and the Severo Ochoa Centres of Excellence Program under grant SEV-2015-0496 and by the Generalitat de Catalunya under grant no. 2017 SGR 1506. X.C. acknowledges financial support by the Ministerio de Economía, Industria y Competitividad under grant TEC2015-67462-C2-1-R (MINECO/FEDER), the Ministerio de Ciencia, Innovación y Universidades under Grant No. RTI2018-097876-B-C21 (MCIU/AEI/FEDER, UE), and the EU Horizon2020 research and innovation program under grant No. GrapheneCore2 785219. M. L. S. is funded through a Juan de la Cierva fellowship. J.M.-S. acknowledges support through the Clarín Programme from the Government of the Principality of Asturias and a Marie Curie-COFUND European grant (PA-18-ACB17-29). R.T. Acknowledges support by European Union's Horizon 2020 research and innovation programme (SPQRel grant agreement no. 679183)

## Author contributions

M.D.L. and I.Z. conceived the experiment. J.M.-S. and R.T. prepared and characterized the samples by AFM. M.D.L. performed the Raman measurements and analyzed the results. X.C., M.L.S., and R.R. performed the theoretical calculations. M.D.L. wrote the manuscript. I.Z. supervised the project. All authors discussed the results and commented on the manuscript.

# Supporting Information

## New insights in the lattice dynamics of monolayers, bilayers, and trilayers of $WSe_2$ and unambiguous determination of few-layer-flakes' thickness

Marta De Luca[1], Xavier Cartoixà[2], Javier Martín-Sánchez[3], Miquel López-Suárez[4], Rinaldo Trotta[5], Riccardo Rurali[4], and Ilaria Zardo[1]

[1]Departement Physik, Universität Basel, 4056 Basel, Switzerland
[2]Departament d'Enginyeria Electrònica, Universitat Autònoma de Barcelona, 08193 Bellaterra, Barcelona, Spain
[3]University of Oviedo, Edif. Severo Ochoa c/ Dr Fernando Bonguera, 33006 Oviedo, Spain
[4]Institut de Ciència de Materials de Barcelona (ICMAB–CSIC), Campus de Bellaterra, 08193 Bellaterra, Barcelona, Spain
[5]Departement of Physics, Sapienza University of Rome, P.le A. Moro 2, 00199 Rome, Italy

*Authors for correspondence, E-mail: marta.deluca@unibas.ch; ilaria.zardo@unibas.ch

## 1. WSe2 samples: preparation and characterization

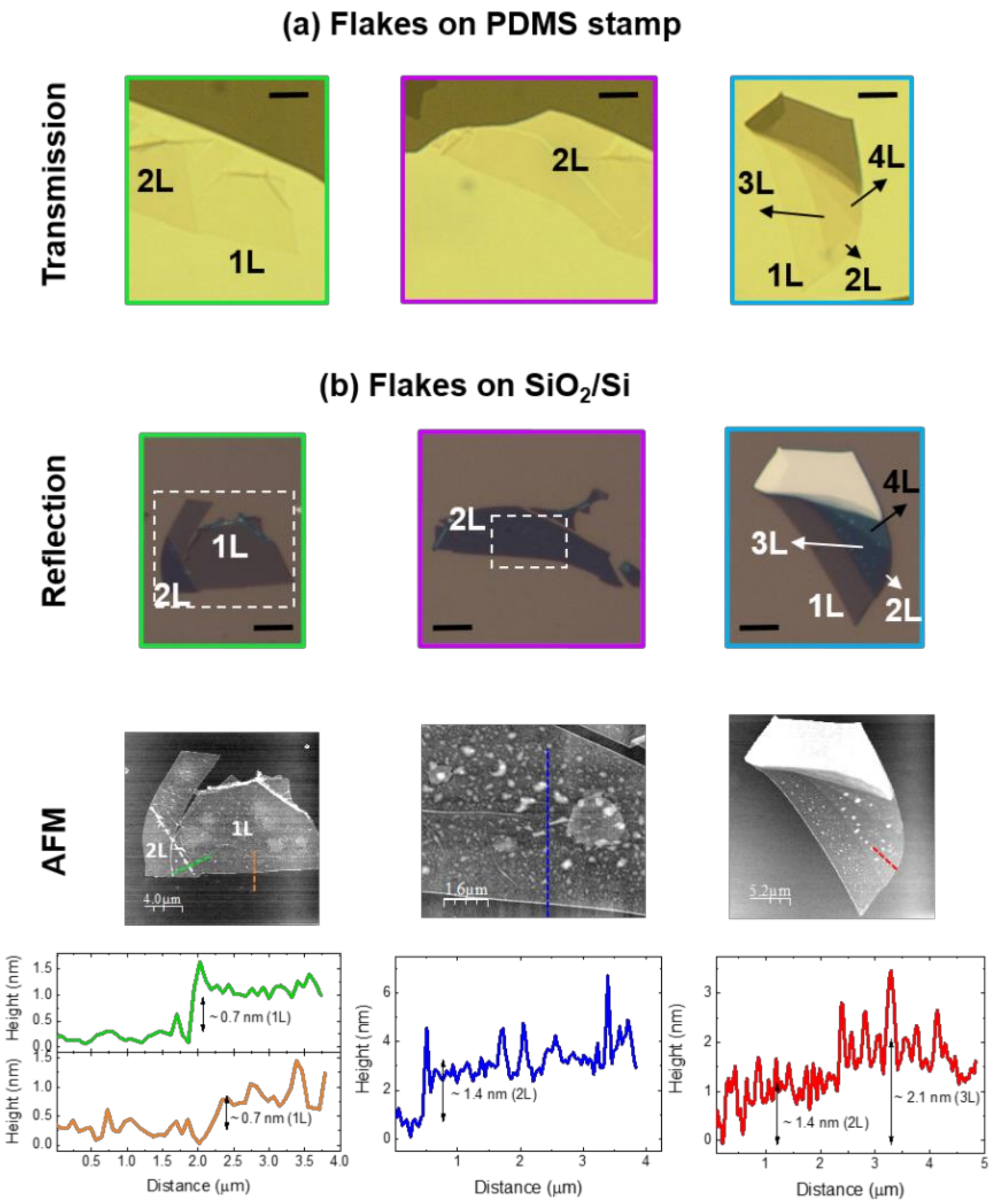

**Figure S1.** (a) Flakes obtained after exfoliating a WSe2 crystal on a PDMS stamp. The flakes on the stamp are localized by measuring optical transmission. With these images, it is possible to distinguish between monolayers (1L), bilayers (2L), and trilayers (3L) quite reliably (the exact determination of the number of layers can also be verified by Raman spectroscopy and atomic force microscopy, AFM). (b) The flakes are deterministically transferred from the selected regions on the SiO2/Si substrate where all the Raman measurements were performed by using a micro-manipulator and then imaged in reflection. In the whole paper, the '1L' sample is the one displayed in the first column, the '2L' sample is the one displayed in the second column, and the '3L' sample is the 3L is the one indicated in the flake displayed in the third column. All the scale bars correspond to 5 μm. (c) Atomic force microscope images (top panels) and profiles (bottom panels) of the 1L, 2L, and 3L selected for the Raman investigation. The AFM analysis confirm the layer thickness that we had determined based on the transmission measurements. Some regions in the 1L and 2L samples appear damaged because a high-power laser was focused there to exfoliate layers for another experiment. However, the Raman measurements of this work were performed far from these damaged regions. Optical images do not show damaged regions because they were acquired before laser exposition.

## 2. Calculated phonon dispersion along ΓM, MK, and ΓK for the 1L, 2L, and 3L

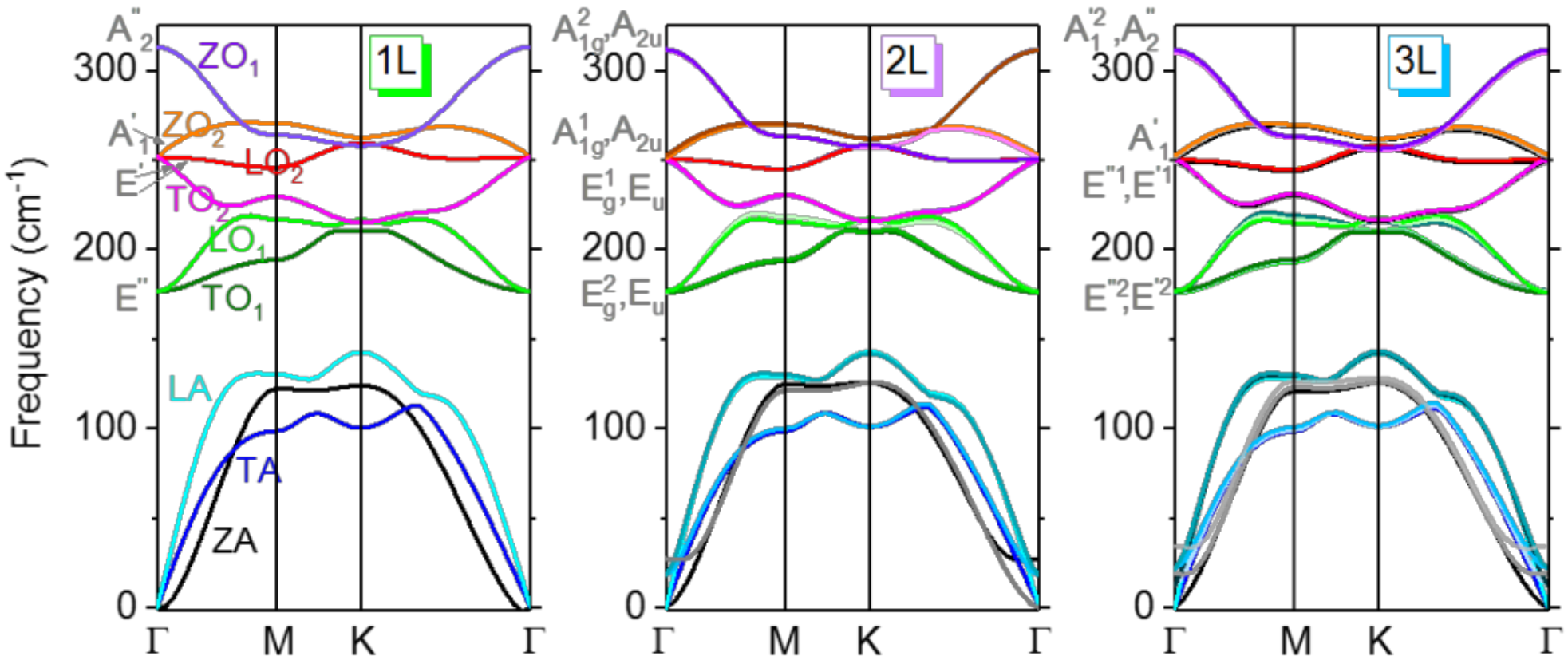

**Figure S2.** Phonon dispersions of the $WSe_2$ monolayer, bilayer, and trilayer (from left to right) calculated by DFPT. For clarity reasons, branches' labels (i.e. ZA, TA, LA, $TO_{1,2}$, $LO_{1,2}$, $ZO_{1,2}$) are given only for the 1L but they apply to all layers, as suggested by the color code. The acoustic branches (ZA, TA, LA) are labeled only with the name of the branch, while for the optical branches, also the symmetry that the phonon modes take at Γ is given (for 1L, 2L, 3L, see grey labels). For identifying those symmetries, it is necessary to consider the symmetry of the unit cell, which is detailed in the main text. In summary, in the 1L the six optical modes at Γ correspond to the irreducible representations: E" ($LO_1$ and $TO_1$), E' ($LO_2$ and $TO_2$), $A_2$" ($ZO_1$) and $A_1$' ($ZO_2$). The positive integer used as a superscript of the mode notation is used to distinguish modes with the same symmetry (*e.g.*, $E_{1g}$ and $E_{2g}$ in the 2L), and is generally applicable to other modes. All modes (Raman and IR active) are indicated. Modes with the subscript u are either silent or only IR active [1]. The values of the phonon frequencies at the three high-symmetry points are given in tables S1, S2, and S3 in the SI3.

In the 1L there are three atoms per unit cell, so there are nine phonon branches. When going from the 1L to the 2L, the number of atoms in the unit cell doubles, and so does the number of the phonon branches. Each of the nine normal vibrational modes in 1L will split into the corresponding two modes in the 2L. For example, E' (R + IR active) in 1L splits into $E_{1g}$ (R) and $E_u$ (IR) in 2L. Similar reasoning applies to other modes and to thicker layers.

## 3. Calculated phonon frequencies at high-symmetry points of the Brillouin zone

| Γ point | | | |
|---|---|---|---|
| **Mode #** | **1L** | **2L** | **3L** |
| 1 | 0 (ZA) | 0 | 0 |
| 2 | 0 (TA) | 0 | 0 |
| 3 | 0 (LA) | 0 | 0 |
| 4 | 176.17 ($TO_1$) | 17.74 | 12.39 |
| 5 | 176.17 ($LO_1$) | 17.74 | 12.39 |
| 6 | 250.76 ($TO_2$) | 27.34 | 19.26 |
| 7 | 250.76 ($LO_2$) | 175.54 | 21.63 |
| 8 | 251.31 ($ZO_2$) | 175.54 | 21.63 |
| 9 | 312.32 ($ZO_1$) | 176.43 | 33.21 |
| 10 | | 176.43 | 175.27 |
| 11 | | 249.57 | 175.27 |
| 12 | | 249.57 | 175.97 |
| 13 | | 249.68 | 175.97 |
| 14 | | 249.68 | 176.51 |
| 15 | | 250.41 | 176.51 |
| 16 | | 251.79 | 248.57 |
| 17 | | 310.50 | 248.57 |
| 18 | | 311.04 | 249.60 |
| 19 | | | 249.60 |
| 20 | | | 249.61 |
| 21 | | | 249.61 |
| 22 | | | 250.08 |
| 23 | | | 251.07 |
| 24 | | | 252.01 |
| 25 | | | 309.19 |
| 26 | | | 310.71 |
| 27 | | | 310.81 |

**Table S1:** Frequencies (in $cm^{-1}$) of the phonon modes calculated at the Γ point of the $WSe_2$ monolayer, bilayer, and trilayer by DFPT. The total number of branches is 9 for the 1L, 18 for the 2L, and 27 for the 3L (see the phonon branches in Figure S2), as expected based on the number of atoms per unit cell (3, 6, and 9, respectively). For clarity, the corresponding branch labels are given only for the 1L.

| M point | | | |
|---|---|---|---|
| **Mode #** | **1L** | **2L** | **3L** |
| 1 | 98.4 (TA) | 98.2 | 98.1 |
| 2 | 122.0 (ZA) | 99.8 | 98.9 |
| 3 | 130.1 (LA) | 121.1 | 100.5 |
| 4 | 193.7 ($TO_1$) | 124.5 | 120.7 |
| 5 | 216.2 ($LO_1$) | 128.4 | 122.7 |
| 6 | 229.1 ($TO_2$) | 129.9 | 125.6 |
| 7 | 245.2 ($LO_2$) | 192.9 | 127.6 |
| 8 | 263.3 ($ZO_1$) | 193.9 | 129.2 |
| 9 | 270.0 ($ZO_2$) | 214.7 | 129.9 |
| 10 | | 217.7 | 192.5 |
| 11 | | 229.8 | 193.3 |
| 12 | | 230.0 | 194.0 |
| 13 | | 244.1 | 214.2 |
| 14 | | 244. 5 | 216.1 |
| 15 | | 262.5 | 218.4 |
| 16 | | 262.6 | 229.7 |
| 17 | | 269.1 | 229.9 |
| 18 | | 269.4 | 230.7 |
| 19 | | | 243.4 |
| 20 | | | 244.3 |
| 21 | | | 244.3 |
| 22 | | | 261.9 |
| 23 | | | 262.5 |
| 24 | | | 262.5 |
| 25 | | | 268.6 |
| 26 | | | 269.2 |
| 27 | | | 269.3 |

**Table S2:** Same as table S1 for the M point.

| K point | | | |
|---|---|---|---|
| **Mode #** | **1L** | **2L** | **3L** |
| 1 | 99.9 (TA) | 100.7 | 100.7 |
| 2 | 123.7 (ZA) | 100.7 | 100.7 |
| 3 | 142.4 (LA) | 125.4 | 101.5 |
| 4 | 209.6 ($TO_1$) | 125.4 | 125.3 |
| 5 | 214.4 ($TO_2$) | 141.8 | 125.4 |
| 6 | 216.2 ($LO_1$) | 142.9 | 127.2 |
| 7 | 257.3 ($ZO_1$) | 209.5 | 141.6 |
| 8 | 258.9 ($LO_2$) | 209.5 | 142.3 |
| 9 | 261.9 ($ZO_2$) | 215.2 | 143.1 |
| 10 | | 215.2 | 209.3 |
| 11 | | 215.7 | 209.5 |
| 12 | | 216.9 | 209.7 |
| 13 | | 255.8 | 214.8 |
| 14 | | 255.8 | 215.3 |
| 15 | | 257.2 | 215.5 |
| 16 | | 258.0 | 215.9 |
| 17 | | 261.2 | 216.2 |
| 18 | | 261.2 | 217.1 |
| 19 | | | 254.4 |
| 20 | | | 255.8 |
| 21 | | | 255.8 |
| 22 | | | 256.1 |
| 23 | | | 257.5 |
| 24 | | | 257.7 |
| 25 | | | 260.7 |
| 26 | | | 261.2 |
| 27 | | | 261.2 |

**Table S3:** Same as table S1 for the K point.

## 4. Thickness-dependence of the first-order phonon modes at ~250 $cm^{-1}$

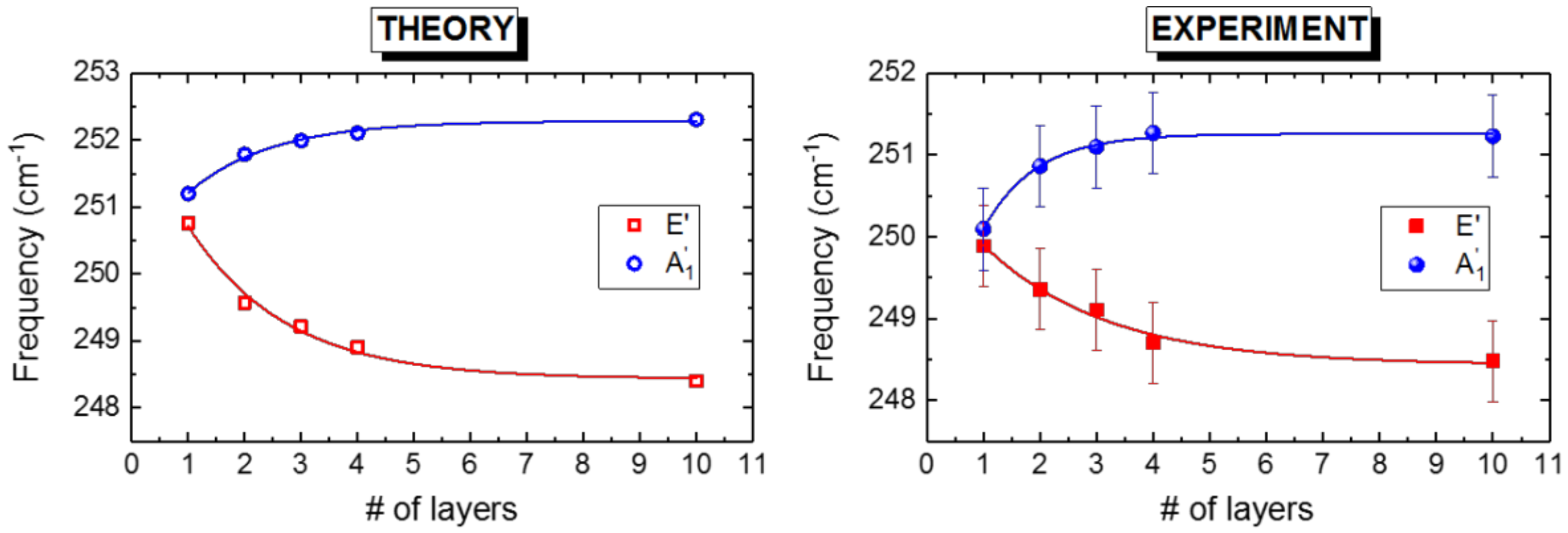

**Figure S3:** Calculated (left panel) and measured (right panel) frequencies of the first-order phonon modes at ~250 $cm^{-1}$ for increasing number of layers. Solid lines are exponential fits to the data. Modes' labels are given using the symmetry of the 1L for simplicity. Notice that the theory scale is extended by 1 $cm^{-1}$ more than the experimental scale to allow to plot the $A'_1$ modes (the general overestimation of the theory frequencies, due to the fact that DFPT calculations are performed a 0 K, is discussed in the methods, section 4.3). Experimental data are relative to the experiments performed with 633 nm excitation wavelength (displayed in Figure 2 in the main text), as longer wavelengths allow for a higher spectra resolution. The experimental and theoretical data corresponding to 10L are actually relative to the bulk crystal, which is indicated as 10L for representation purposes. Theoretical and experimental data were derived by fitting the relative spectra: while the frequency of the E' phonon modes was extracted by fitting the data/calculations in the xy geometry, the frequency of the $A'_1$ modes was derived by averaging the frequency obtained in both xx and xy geometry, because in the xx geometry we observe both the E' and the $A'_1$ phonon modes (see Figure 2). In the case of the 3L, in which the theory predicts a splitting of the low-frequency mode (see discussion of Figure 2 and ref. [2]**Error! Reference source not found.**), the doublet was fitted with only one curve, in order to average out the effect of the splitting and obtain results directly comparable to our experimental data.

Our theory predicts a shift for the low-frequency mode slightly larger than that of the high-frequency mode, in agreement with ref. [1] when LDA is used. However, in ref. [1], the two modes were not experimentally distinguished. Our experimental data follow our predicted trend, although less clearly than in the theory. A similar result is obtained in the experiment in ref. [3], where, however, this behavior is not commented. In the rest of the literature, the shifts of the E'-mode was found to be equal or slightly smaller than that of the $A'_1$-mode [2][4].

## 5. Effects of polarizers' angle on selection rules

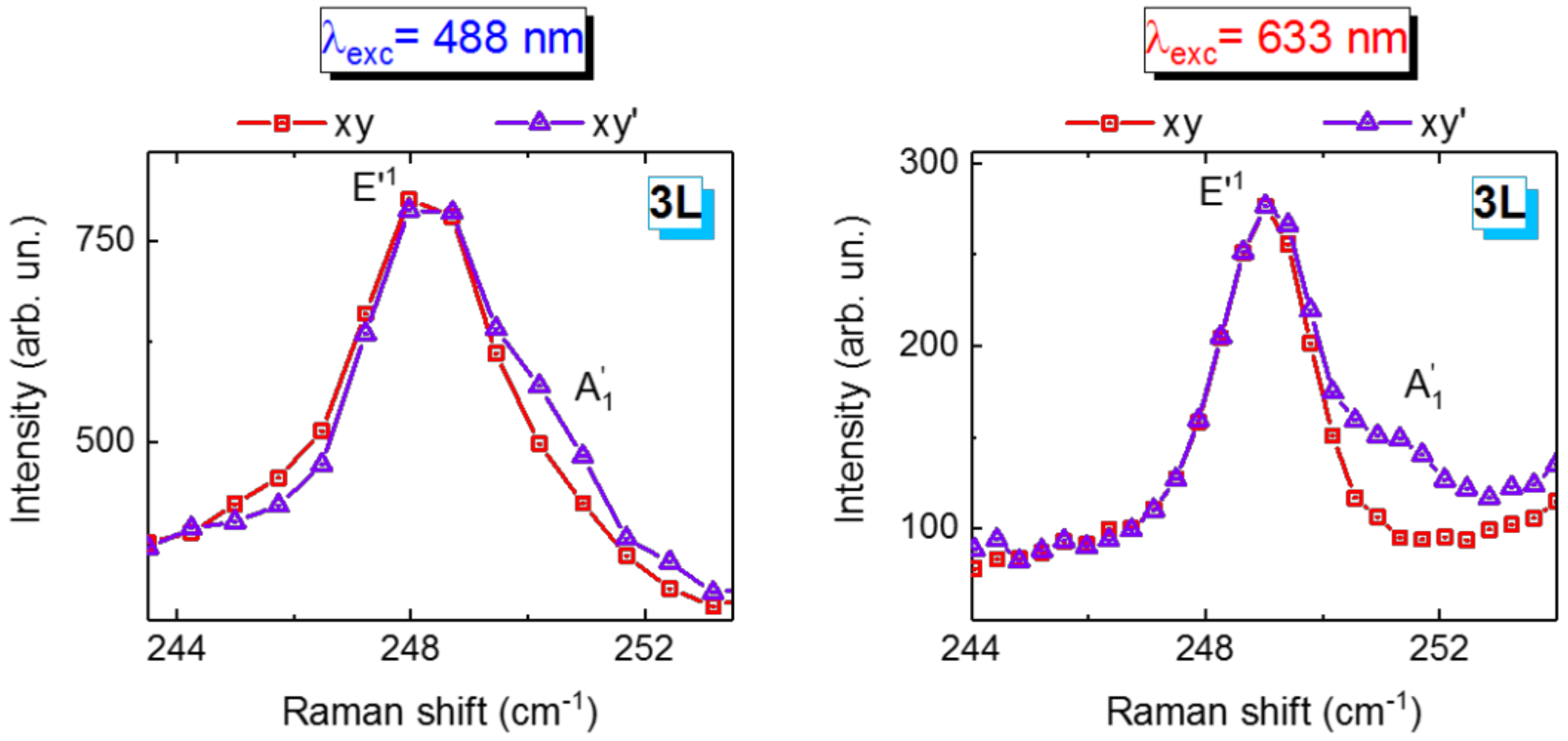

**Figure S4:** Experimental Raman spectra measured with excitation wavelengths of 488 nm (left panel) and 633 nm (right panel) on the 3L sample in the xy (red lines and squares) and xy' degrees (purple lines and triangles) scattering geometries, where y'=y+10°. Since the typical error on commercially available polarizers is about 5 degrees, and since two polarizers are used in polarizers Raman scattering experiments, it is not unlikely to have a 10° error on the relative polarization between excitation and detection light vectors. Clearly, such a small mistake in the angle of the polarization, allows forbidden modes (in this case the $A'_1$) to appear.

## 6. Polarization-dependence, wavelength-dependence, and thickness-dependence of the '2LA' band at ~260 cm-1

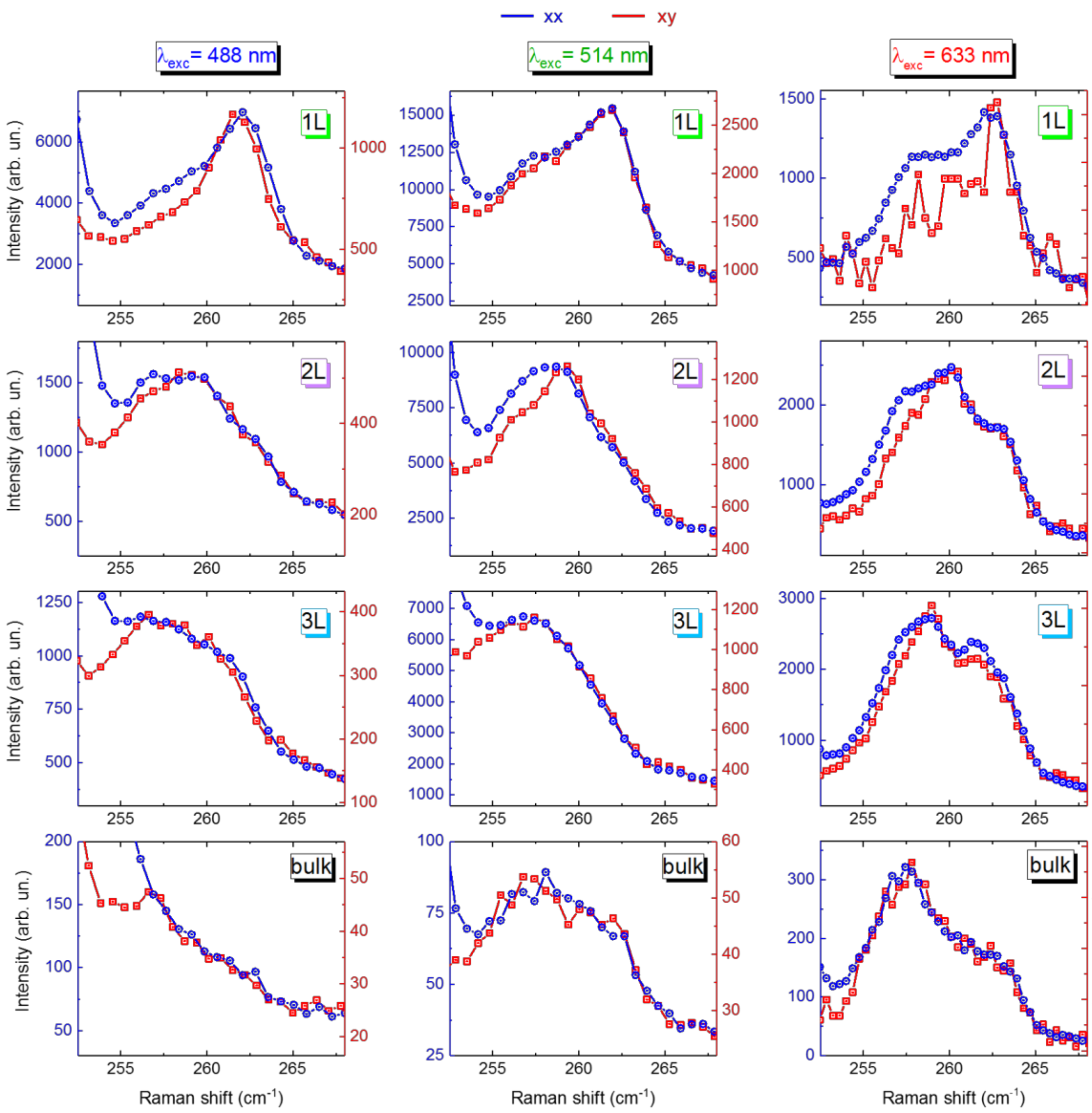

**Figure S5:** Raman spectra of the 1L, 2L, 3L, and bulk samples in the xx (blue lines and circles) and xy (red lines and squares) scattering geometries magnified in the region of the first- and second-order phonon modes at ~ 260 cm-1. Spectra were collected with the 488, 514, and 633 nm excitation wavelengths (from left to right columns, respectively). For each couple of spectra (xx and xy), intensity scales were chosen in order to approximately align the background signal on the high-frequency side as well as the maxima.

Due to the large number of phonon modes contributing to this band, and the fact that selection rules for those modes have not been calculated, a detailed analysis taking into account the polarization-dependence of the modes would be arbitrary.

7. **Thickness-dependence of the first- and second-order phonon modes composing the ‘2LA’ band at ~260 cm-1**

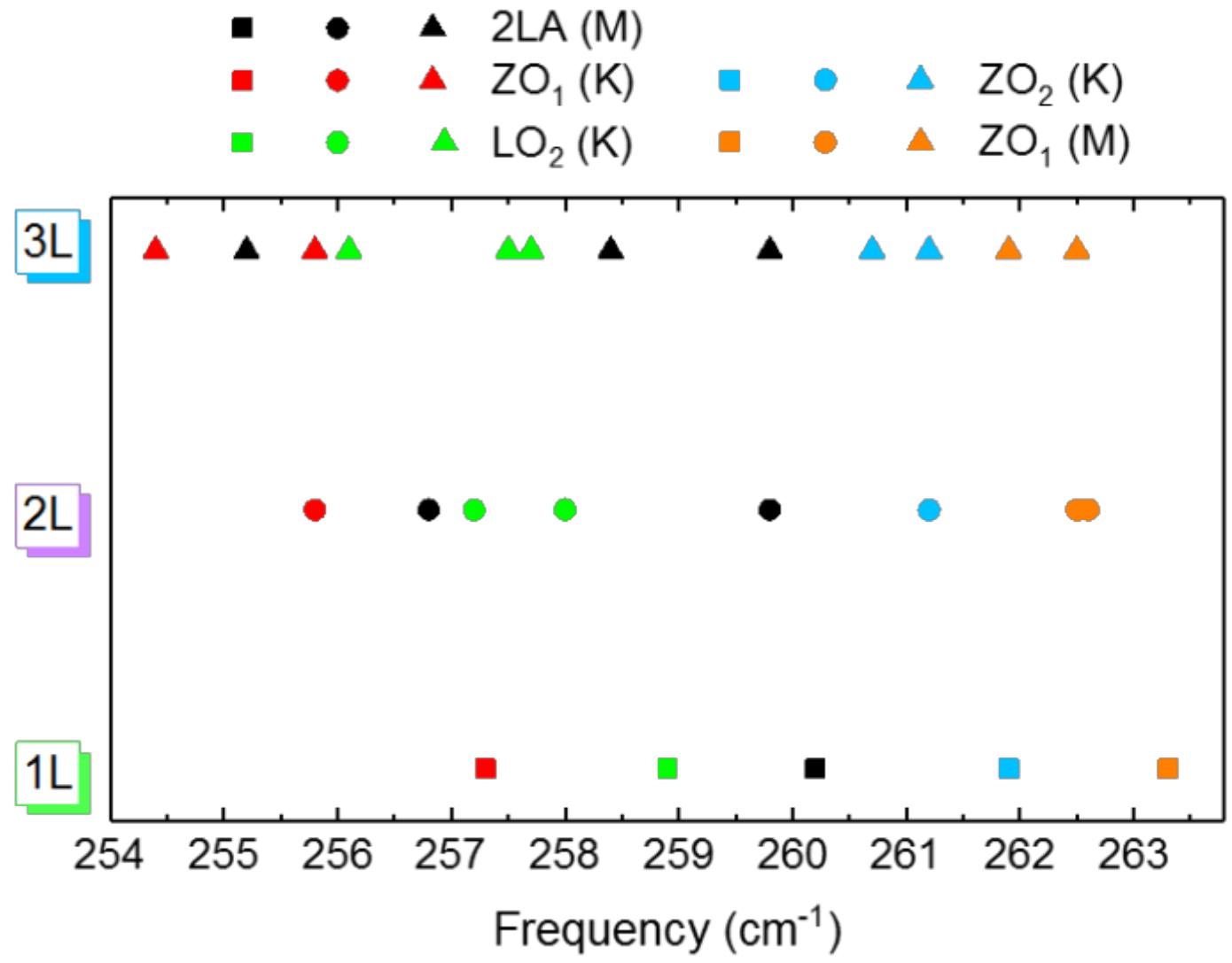

**Figure S6:** Calculated frequencies of the phonon combination (black symbols) and of single phonons (colored symbols) for the 1L (squares), 2L (circles), and 3L (triangles) samples that are likely to provide the most important contributions to the broad 254-264 cm-1 band, in literature commonly interpreted as ‘2LA’ band.

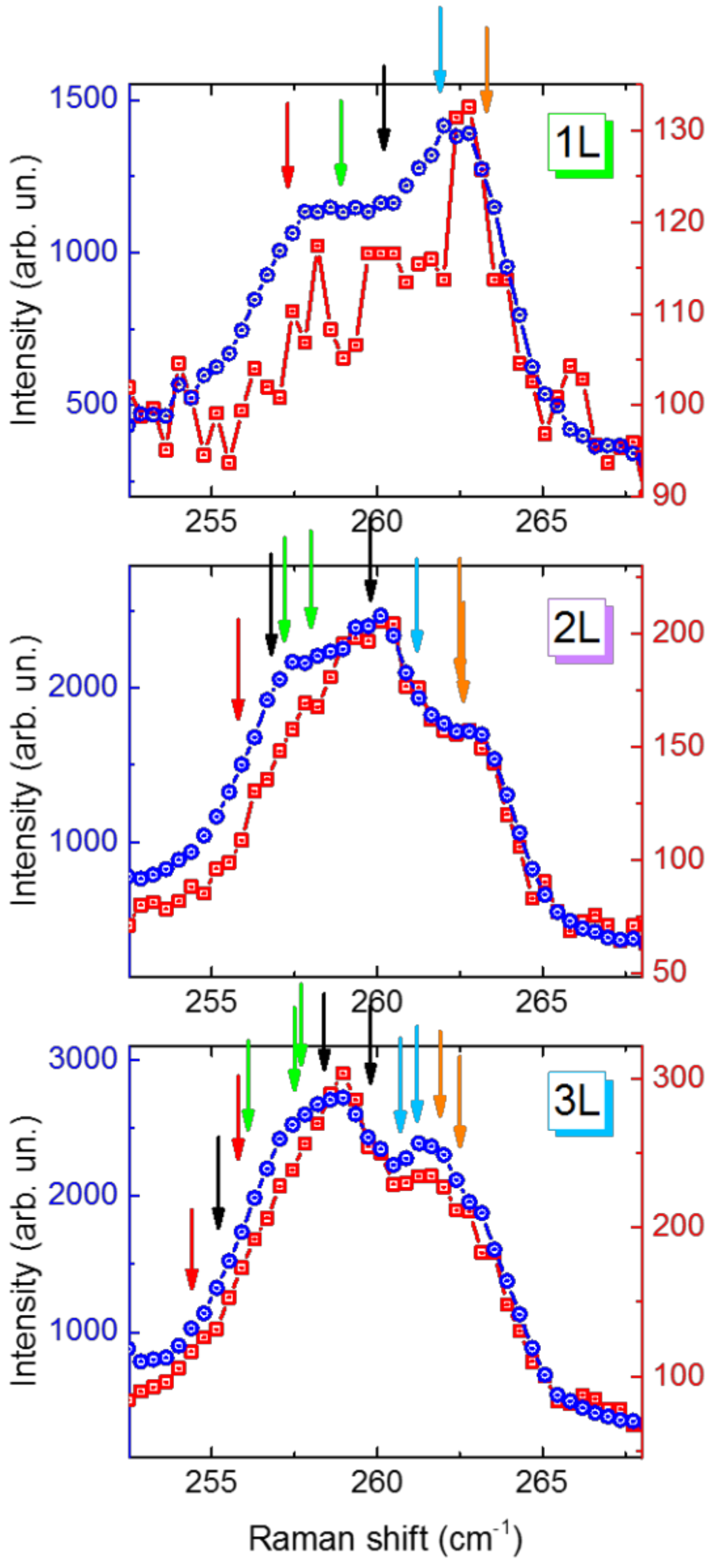

**Figure S7:** Raman spectra of the 1L, 2L, and 3L in the xx (blue lines and circles) and xy (red lines and squares) geometries magnified in the region of the band at ~ 260 cm-1. Intensity scales were chosen to approximately align the background signal on the high-frequency side as well as the maxima, for clarity. Spectra were collected with 633 nm wavelength (same spectra, only in xx, are shown in Figure 3, third column, of the main text). The arrows mark the frequencies displayed in Figure S6 and follow the same color code. In the 1L, nearly all the contributions were identified in the deconvolution (this is shown in Figure S8). In the 2L and 3L, including in the fitting procedure the expected large number of contributions would have involved a non-negligible degree of arbitrariness. However, the gradual downshift of the contributions when going from the 1L to the 3L predicted from theory (Fig. S6) is observed also in the experimental data and it is likely to be responsible for the change in the spectral weight of this band. It is not possible to properly interpret the different enhancement of the different contributions

of this broad band when going from 1L to 3L, because the two existing resonant Raman studies of this band do not agree with each other [2][5]. However, since the clear change in the spectral weight is detectable with all the excitation wavelengths in the visible range (see Figure 3 in the main text), it can provide a universal method to distinguish 1L, 2L, and 3L with Raman spectroscopy.

8. **Second-order phonon modes in the Raman spectra of the monolayer (spectral range: 215-273 cm-1)**

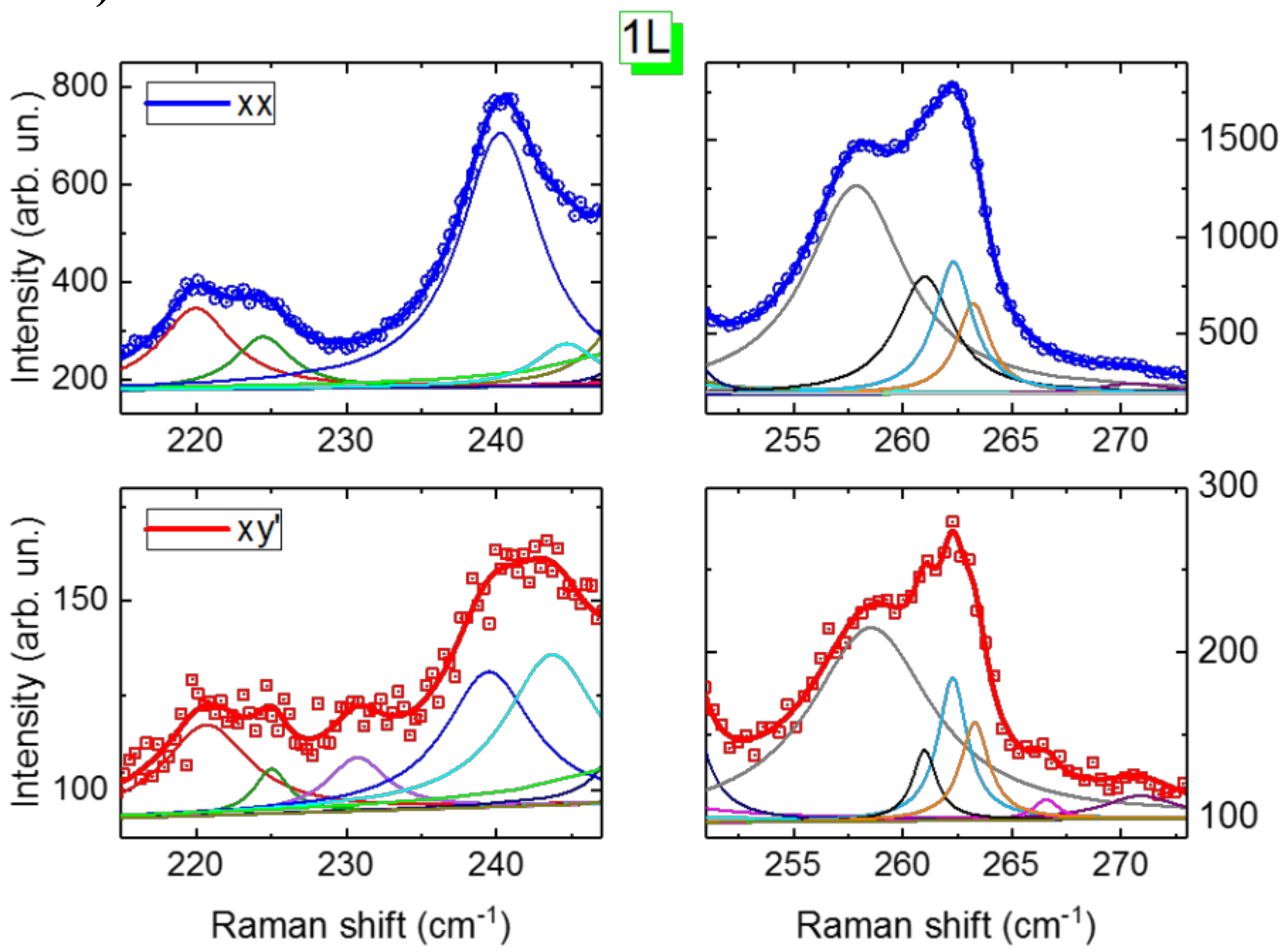

**Figure S8:** Experimental Raman spectra acquired with the 633 nm wavelength on the 1L sample in the xx (blue lines and circles) and xy' [y'=y+10°] (red lines and squares) scattering geometries. The detection polarization in the xy' geometry was selected to be slightly different from the conventional xy geometry in order to have a non-zero signal from otherwise forbidden modes (see 1L spectra in xy geometry in Figure S5 and S7) and be able to perform a quantitative analysis of the peaks. The spectra are divided into two different frequency ranges (215-247 on the left and 251-273 cm-1on the right) for clarity reasons. The range where the first-order phonon mode at 250 cm-1 is located is not shown here, as it is extensively shown and discussed in the main text. Thick solid lines are fits to the data and thin colored lines are the single lorentzian components (a same color in xx and xy spectra indicate a same phonon mode). For the modes that are discussed also in Figures S6 and S7 the same color of Figures S6 and S7 code is used. For the broad peak interpreted as a convolution of $ZO_1$(K) and $LO_2$(K) we have used the grey color.

The frequencies extracted by the fits and located in the range 239-265 cm-1 are given in table 1 in the main text. The peaks centered at ~ 220, 224, 231, 267, and 270 cm-1 are very clearly visible also in the data taken with the 488 nm excitation wavelength (on which most of table 1 is based), and we obtain a very good agreement between the frequencies determined in the 488 and 633 nm Raman spectra. Therefore, for those peaks, in table 1 we display the frequencies extracted by the 488 nm Raman data for consistency. The modes at ~ 231 and 267 cm-1 are likely to come from a combination of phonons associated to in-plane vibrations because they are observed only in xy geometry.

## 9. Calculated two-phonons density of states

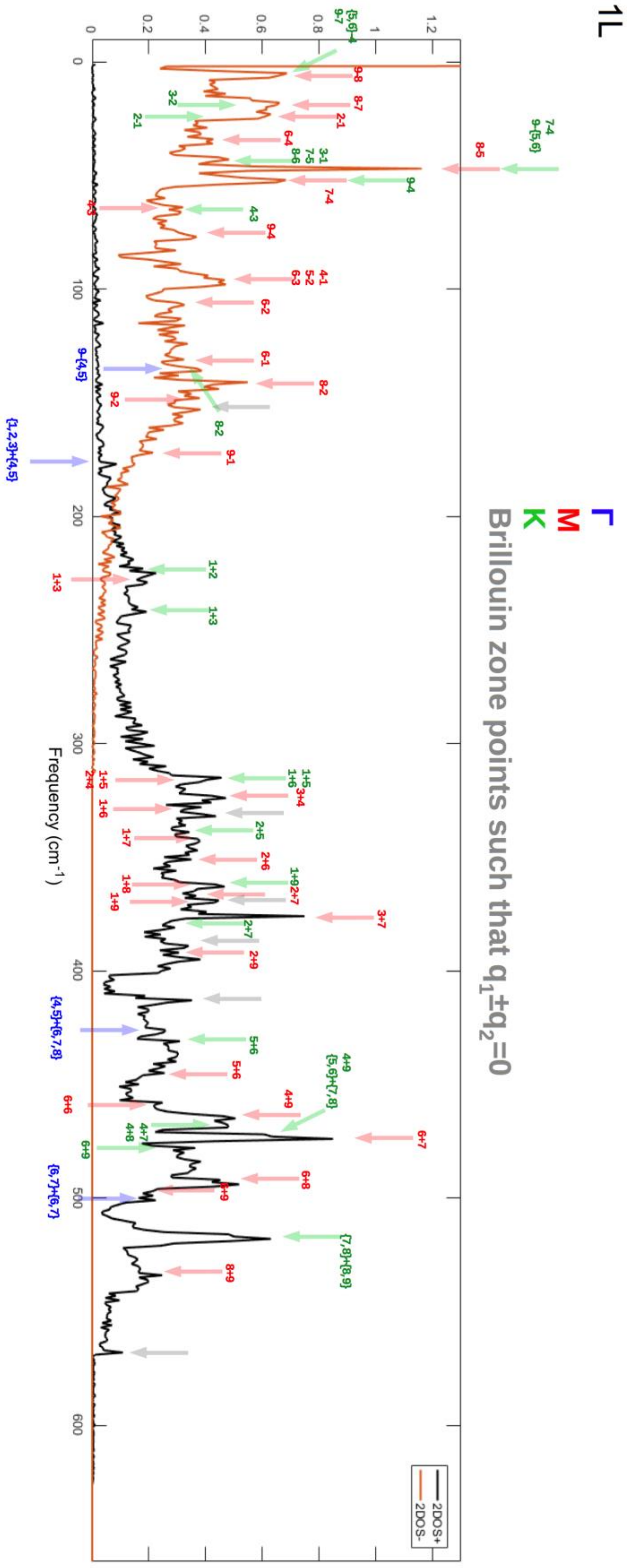

2L

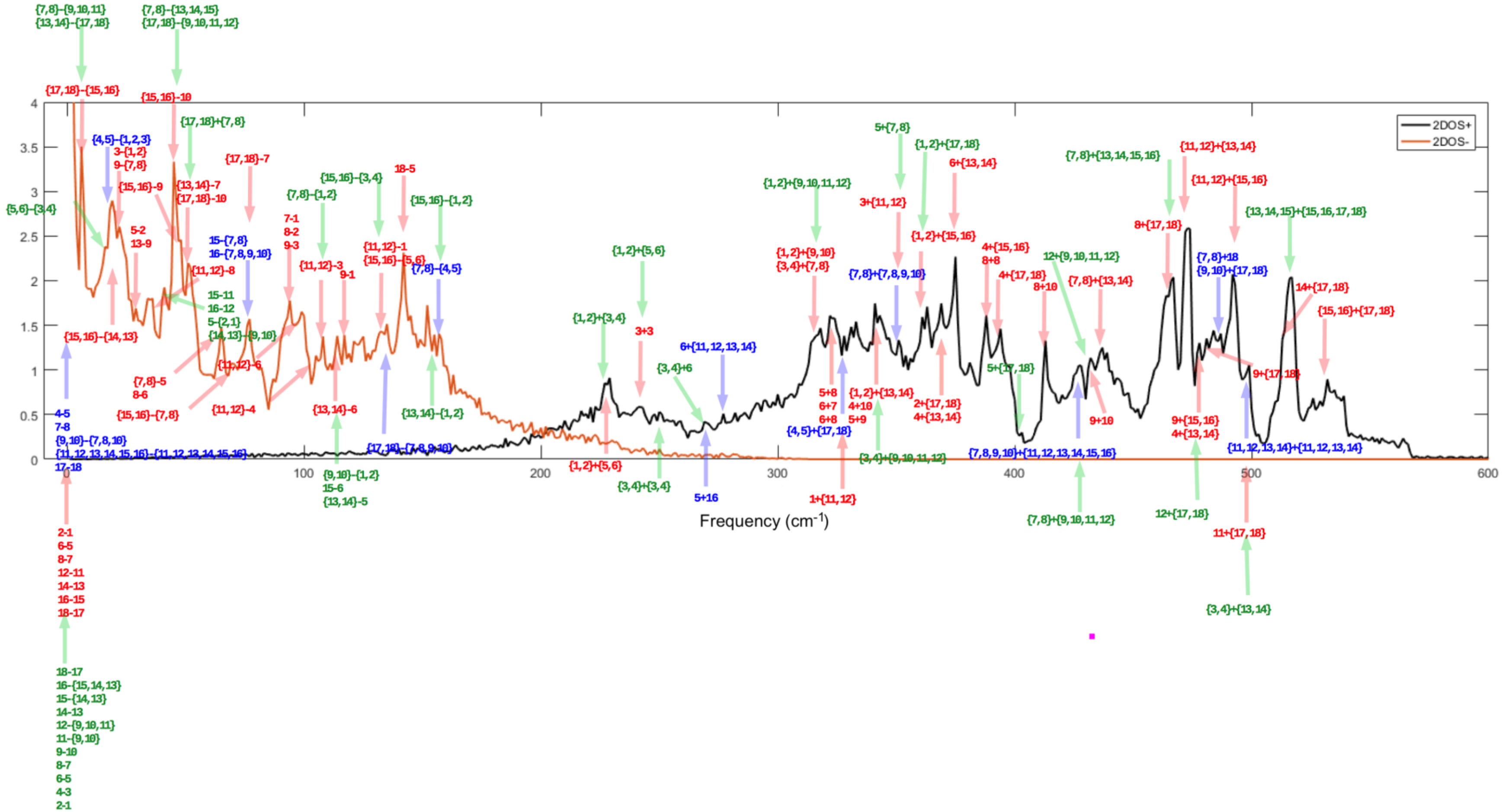

3L

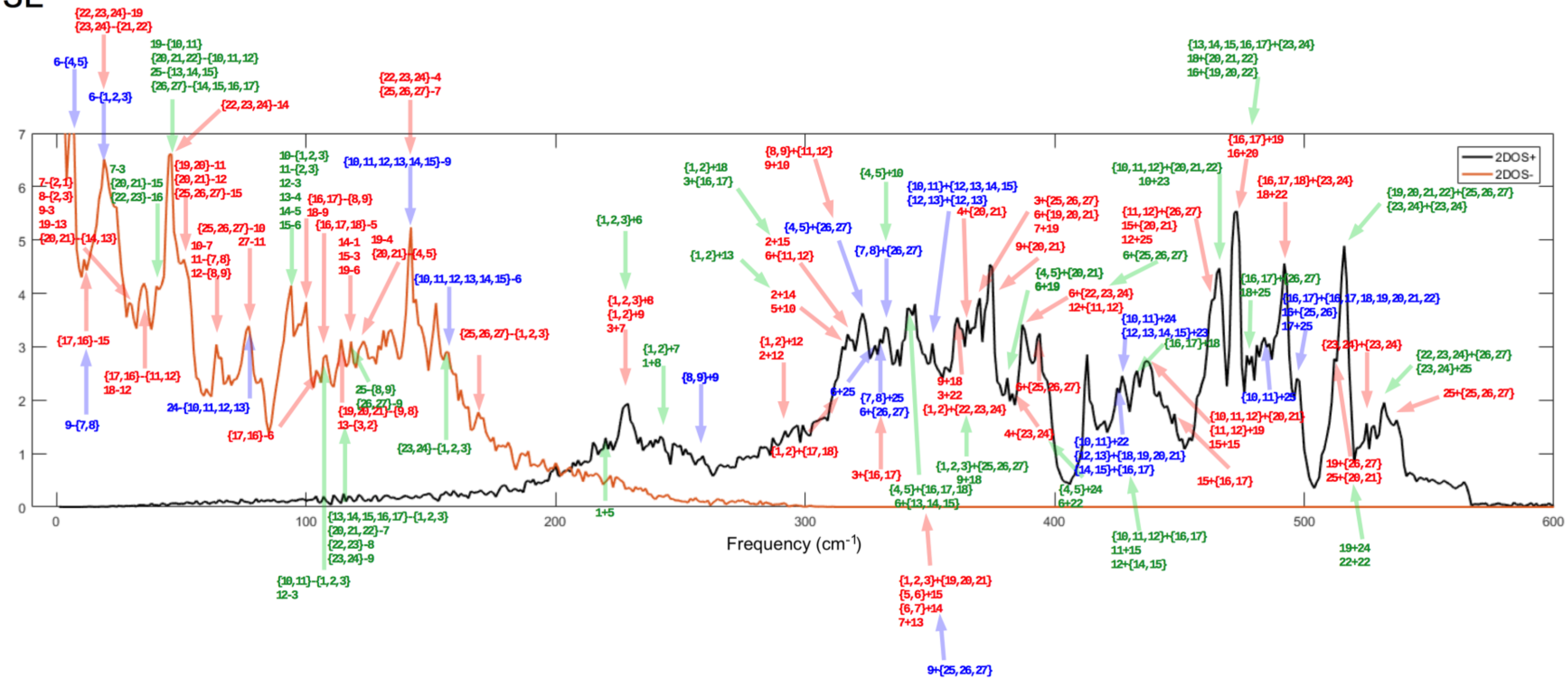

**Figures S9:** Calculated two-phonon density of states (2DOS+ in black and 2DOS- in orange, corresponding to absorption and emission processes, respectively) for the 1L, 2L, and 3L. Arrows point to the most intense/clear peaks in the curves. The color of the arrows indicates the wavevector of the phonons contributing to each peak (see the legend), and the number in parenthesis labels the phonon modes according to the row numbers given in tables S1, S2, and S3.

## 10. Thickness-dependence of the phonon modes at ~220, 224, and 242 cm-1

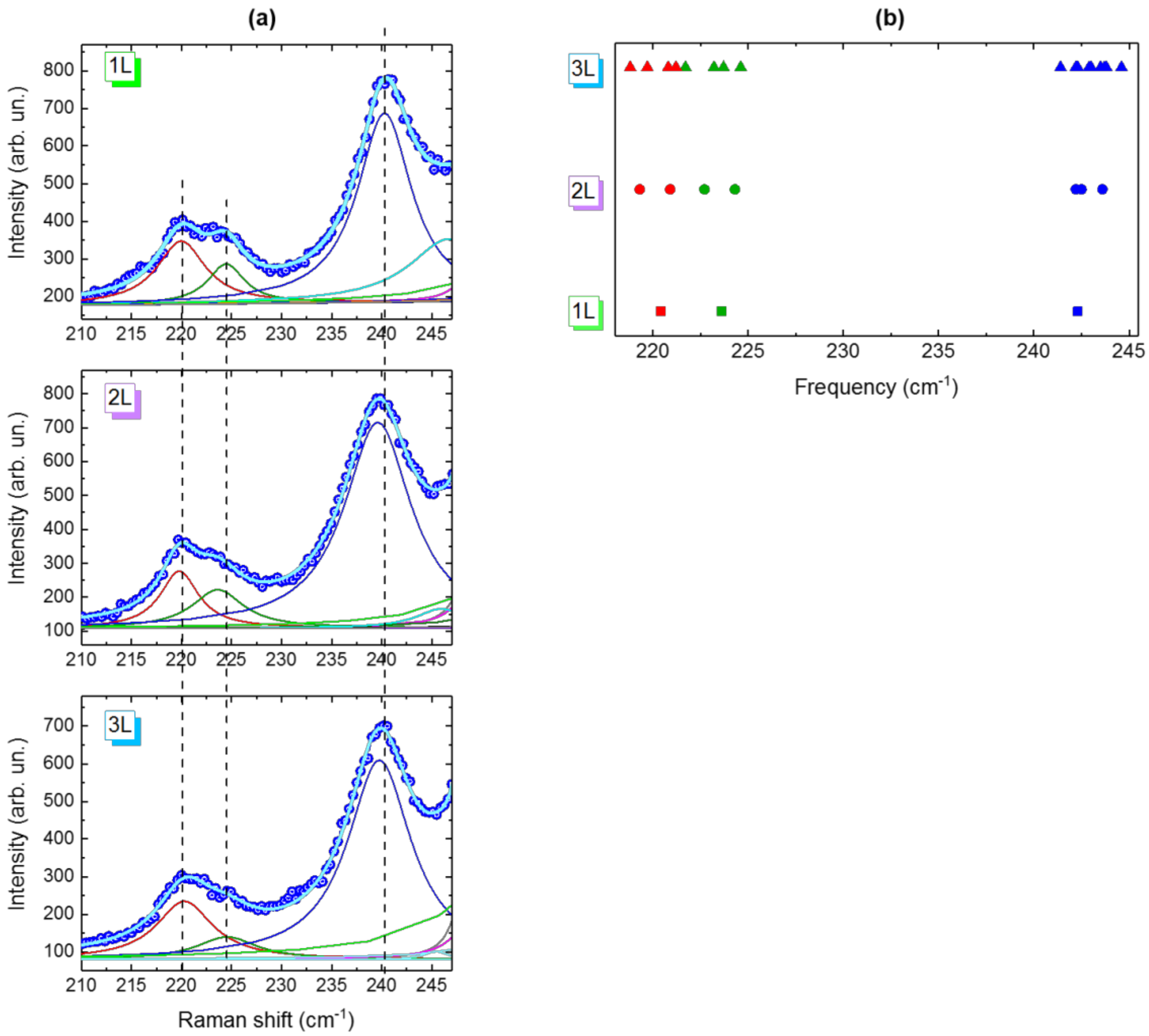

**Figure S10:** (a) Experimental Raman spectra acquired with the 633 nm wavelength on the 1L, 2L, and 3L samples in the xx geometry in the range 210-246 cm-1. Thick solid lines are fits to the data, thin colored lines are the single lorentzian components. For the 1L, the deconvolution is the same displayed in Figure S8. Dashed lines mark the position of the peaks in the 1L. (b) Calculated frequencies of the phonon combinations for the 1L (squares), 2L (circles), and 3L (triangles) that are likely to account for the experimental peaks observed in the 220-245 cm-1 range. For the 1L, the interpretation of the three modes at ~ 220, 224, and 242 cm-1, visible also with the 488 nm excitation wavelength, is given in table 1. For example, the peak at 224 cm-1 was assigned to TA(K) + ZA(K). Clearly, in the 2L and 3L there are more possible contributions resulting in a same frequency. In the 2L and 3L samples, including in the fitting procedure the expected large number of contributions would involve a non-negligible degree of arbitrariness. However, for each of the three main contributions, the average theoretical frequency approximately matches the experimental data.